\documentclass[]{aastex631}
\usepackage[utf8]{inputenc}
\usepackage{todonotes}
\usepackage{comment}

\newcommand{\kms}{\ensuremath{\textrm{km}\,\textrm{s}^{-1}}}
\newcommand{\kmss}{$\rm km~s^{-1}$ }

\newcommand\ha{{H$\alpha$}}

\newcommand\hbeta{{H$\beta$}}
\newcommand\VEL{\:{\rm km\:s^{-1}}}

\begin{document}
\title{Observed characteristics of M31 SNR}

\title{Spectroscopy of Supernova Remnants and Candidates in M31}

\correspondingauthor{Nelson Caldwell}
\email{caldwell@cfa.harvard.edu}

\author[0000-0003-2352-3202]{Nelson Caldwell}
\affiliation{Center for Astrophysics, Harvard \& Smithsonian, 60 Garden Street, Cambridge, MA 02138, USA}
\author[0000-0002-7868-1622]{John C. Raymond}
\affiliation{Center for Astrophysics, Harvard \& Smithsonian, 60 Garden Street, Cambridge, MA 02138, USA}
\date{Oct 2024}

\author[0000-0002-4134-864X]{Knox S. Long}
\affiliation{Space Telescope Science Institute,
3700 San Martin Dr,
Baltimore MD 21218, USA}
\affiliation{Eureka Scientific, Inc.
2452 Delmer Street, Suite 100,
Oakland, CA 94602-3017}

\author[0000-0003-2713-6744]{Myung Gyoon Lee}
\affiliation{Astronomy Program, Department of Physics and Astronomy, SNUARC, Seoul National University, 1 Gwanak-ro, Gwanak-gu, Seoul
08826, Republic of Korea}

\begin{abstract}

With a star formation rate of order 0.4 M$_\sun $ yr$^{-1}$, M31 should have significant population of supernova remnants (SNRs), and, in fact, 156 SNR and SNR candidates have been suggested by \cite{lee14} by searching for nebulae with elevated [S~II]/\ha\ ratios in narrow band images.   Here we use a combination of low and high resolution optical spectroscopy obtained with Hectospec on the MMT to characterize 152 of these nebulae.  Of these candidates, we find 93 nebulae that have [SII]/\ha\ ratios that exceed 0.4, the traditional ratio used to separate SNRs from HII regions, strongly suggesting that at least these objects are SNRs.  Our high resolution spectroscopy reveals 108 nebulae that have velocity widths in \ha\  (full-width at 20\% peak flux) that exceed 50 $\VEL$, significantly larger than found in HII regions.  There are 72 objects that satisfy both tests.  Here we discuss the spectroscopic  characteristics of all of the objects in our sample, and the likelihood that other objects in the sample of  \cite{lee14} are also SNRs, and we briefly consider confirmation by X-ray, radio and UV observations. We also discuss several new candidates that have been identified serendipitously in the course of examining a large amount of archival Hectospec data.

\end{abstract}

\section{Introduction}

Supernova remnants (SNRs) are the echoes of stars which have exploded and are shaping and enriching the ISM in preparation for the next generation of star formation.  The detectability of a SNR and the age at which it can be detected depend not only on the nature of the SN explosion, but also on the environment into which the SN shock is expanding.  Most SNRs in our Galaxy were first observed as extended non-thermal radio sources. Most SNRs in other galaxies have been identified optically, as emission nebulae with emission line ratios, usually [S~II]$\lambda\lambda$6713,6331/\ha, that distinguish them from HII regions.     In HII regions, most of the \ha\/ emission arises due to recombination of photoionized gas and most sulfur is doubly ionized, leading to observed [SII]/\ha\ line ratios, at least in high surface brightness nebulae, of  typically less than 0.1.  In SNRs, however, most of the emission arises in radiative shocks traversing relatively dense circumstellar or interstellar gas.  The recombining plasma in these radiative shocks contains an extended region with a temperature of 5,000 - 20,000 K where singly ionized sulfur dominates, leading to elevated  [SII]/\ha\ line ratios, with a ratio of 0.4 or greater often being used as clear evidence that a nebula is a SNR. 

There are today about 300  SNRs which have been identified in the Galaxy \citep{green24}.  Although a substantial number of these SNRs have been studied in great detail, uncertainties in distance and the effects of variable line of sight absorption restrict the utility of Galactic SNRs for determining properties of SNRs as a class.  Much less is known  about at least 2000 nebulae, known or suggested to be SNRs in of order 20 external galaxies, but SNRs in an external galaxy are all at the same distance and typically suffer less from the effects of variable absorption. As a result, they provide a better laboratory for studying SNRs as a class \cite[see, e. g.][for a review]{long17}.    

Galaxies within the Local Group are clearly the most important for such studies because sensitivity and spatial resolution considerations imply we can learn more about SNRs in the nearest ones than those that are more distant.  There are approximately 100 SNRs known in the Magellanic Clouds and the majority of these have been detected at radio, IR, optical, and X-ray wavelengths.  There are 220 SNRs  and SNR candidates that have been identified in M33; of these at least 178 have spectroscopically-determined [SII]/\ha\ ratios greater than 0.4, 113 have been detected in X-rays, and 155 have been detected in the radio \citep{long18, white19}  Most of the SNRs detected are  soft X-ray sources and many are known to have non-thermal radio spectra as expected.  

Of all the galaxies in the Local Group, M31, with a SFR of 0.35 - 0.4  M$_\sun $ yr$^{-1}$ \citep{m31_sfr}, most resembles our own, and a substantial population of SNRs should exist.  Not surprisingly, SNR candidates have been identified, beginning with \cite{ddb80} who identified 19 candidates based on interference filter imagery.   \cite{bkc81} subsequently confirmed spectroscopically that 14 of the candidates had [SII]/\ha\ ratios greater than 0.4, and should be classified as bona-fide SNRs.  The procedure of identifying SNR candidates first as nebulae with elevated [SII]/\ha\  ratios in interference filter imagery followed by spectroscopic confirmation is the traditional way SNRs were identified until very recently, when IFU spectroscopy with MUSE  \cite[e.g.][]{cid-fernandes21, long22}  and SITELLE  \citep{vicens-mouret23,duarte-puertas24} has begun to allow one to collapse the process of identification into a single observation for galaxies outside the Local Group. However, for the case of M31, which is too large to consider IFU surveys with existing instrumentation, the latest and most detailed search for SNRs was carried out in a traditional manner by \cite{lee14} (LL14), who identified 156 SNR candidates in the galaxy.  They identified SNRs as relatively isolated, approximately circular nebulae with elevated [SII]/\ha\ ratios in images obtained by \cite{massey}, excluding a number of candidates that had been suggested by others \citep{bw93,mpvlshpt95} using lower quality data, but which failed to meet their criteria.  This number is not too different from that expected:  Indeed, based on the absolute K and B magnitude of M31 and SN rates estimated by \cite{li11}, there should be between 320 and  700 SNRs that are less than 20,000 years old in M31, and one would not expect to detect all of these, since many SNe explode in regions where a previous explosion has evacuated a region, making subsequent  SNRs extremely difficult to detect.

The purpose of this report is to characterize the SNRs and SNR candidates in M31 identified by LL14 spectroscopically.   For this purpose we use low and high resolution spectroscopy  obtained  with Hectospec and Hectochelle on the MMT telescope, including some low resolution spectra obtained by LL14 but not fully analyzed until this work.   At a basic level, we seek to determine which candidates are actually SNRs, using the traditional method of measuring line ratios in low (R=1000) spectra and using the line ratios to establish which emission nebulae are shock-dominated SNRs and which are photoionized HII regions.  However, as was noted recently by   \cite{plwb19} it is also possible to separate HII regions and SNRs kinematically by obtaining spectra with higher spectral resolution than was traditional, and with higher efficiency spectrographs such observations are now practical.  In SNRs,  radiation arises from  material  moving at speeds that are close to that of the shock velocity typically 100-300 $\VEL$ and as a result, though one only sees the projected component of the shock velocity, most SNRs have line profiles that are broader than 50 $\VEL$; HII regions by contrast are typically expanding at velocities of only 10 or 20 $\VEL$ and therefore have line profiles that are narrow. Furthermore, information about the velocities of the emitting material provides critical information about the the nature of the shock interaction in individual SNRs, and thus should be very useful in explaining why SNRs of similar sizes often have very different optical emission properties.  This study covers 140 of the 156 SNRs and SNR candidates identified by LL14.  The distribution of targets with high [S~II]/H$\alpha$ ratios follows the star formation distribution as expected, with few objects in other locations.  

\begin{figure}[ht]
\includegraphics[width=7.0in]{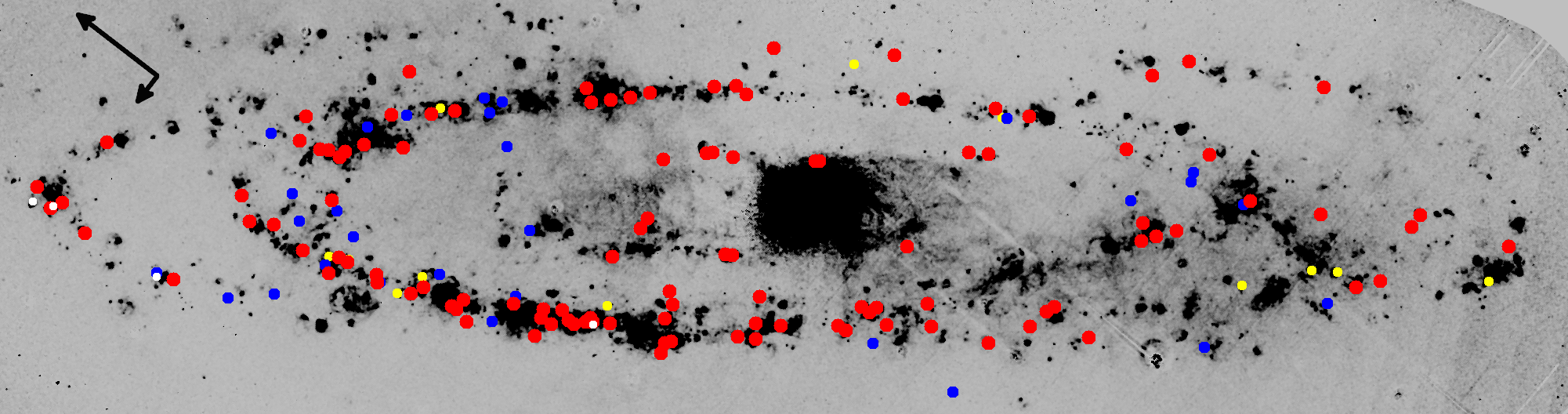} 
\caption{Positions of the LL14 M31 SNR and SNR candidates observed in this study, added to a continuum-subtracted \ha \ image provided by Frank Winkler, which is about 2.4\degr \ wide. Red colors indicate LL14 objects we are certain are SNR, while blue colors show those we are certain are not, most being HII regions. A lone SNR  (LL14-001) is off to the right of the field by about 0.5\degr . Yellow is used for borderline or uncertain cases. The four unobserved LL14 candidates are shown in white. The north arrow is 10\arcmin \  long, or about 2.3 kpc (the east arrow is the shorter one).}
\label{fig:position}
\end{figure}

The remainder of this paper is organized as follows: In Section \ref{data}, we describe the observations and the data that have been reduced for the survey.  We
consider the sample of SNR candidates in \citet{lee14} along with a few serendipitously discovered candidates. Note that we use the nomenclature "LL14-NNN" where NNN is the number in the LL14 paper. In Section \ref{reduction} , we explain how the data were reduced to produce line fluxes and measurements of line broadening.  In Section \ref{criteria} we consider the different methods used to discriminate between SNRs and H II regions.  Section \ref{others} describes some SNR candidates not in the \citet{lee14} list, Section \ref{discuss} presents a discussion, Section \ref{few} briefly describes the complex kinematics of the most interesting observations, and Section \ref{summary} summarizes our results.

\section{Data Description \label{data}}

The spectroscopic data were all taken with the MMT telescope, using both the low-resolution Hectospec fiber-fed instrument and the high resolution Hectochelle spectrograph \citep{Fabricant2005, Szentgyorgyi2011}. The resolutions are 1000 and 32,000, respectively. The low resolution data cover the $\lambda\lambda 3700-9200$ \AA \ region, while the high resolution data cover only the 150 \AA\ region centered on H$\alpha$.  On the sky, the fibers cover 1.5\arcsec \ in diameter.
 

Most of the low-resolution SNR spectra were obtained in 2015 and 2016 at the specific request of Lee \& Lee, but are published here for the first time.
We retrieved them from the CfA archive for this analysis, as well as approximately 15 others taken as part of other M31 emission-line projects for the first author, over an extended period from 2004 to 2023.   A total of 152 of the 156 SNRs and SNR candidates were observed, either at low or high resolution, or both, sometimes multiple times.  Consequentially, the total exposure time for each object ranges from 1400 to 11,000 seconds, with a median value of 5400s. Given the 1.5\arcsec \ diameter fiber, which corresponds to 5.5 pc  at the distance 761$\pm$11 kpc \citep{li21}, the spectra do not provide a global spectrum for the candidates, most of which have sizes between 20-60 pc.  As a result the locations of the fibers are often offset somewhat from the SNR centers listed by LL14, sometimes by as much as 10\arcsec. Nonetheless, careful inspection of narrow-band images shows that the regions observed do correspond to the object whose center was listed in LL14. 

The high resolution MMT SNR  spectra were obtained in 2023 and 2024 as a new program, using fiber positions based on the earlier data. For the most part we used the same positions as the earlier low resolution observations. The dates were 2023-11-06, 2023-11-24, 2023-12-22,and 2024-11-10 with exposure times being between 7200 and 12000 seconds. In some cases, we observed the same SNR at slightly different positions because the object was listed in more than one catalog, with a small offset in position. Specifically, we allowed an offset of up to 5\arcsec \ between the high res and low res data, in a few cases where it was clear the same object was being observed.  A total of 152 LL14 objects were measured in at least one of the two resolution modes, with 135 in the low resolution and 147 in the high. 130 were measured in both.


Because we wanted to compare spectra of the SNR and SNR candidates to spectra of emission nebulae with photoionized gas, we have also retrieved from the CfA archive and reanalyzed low-resolution spectra from previous published M31 projects:  \cite{SCM} on HII regions and PNe, \cite{miko14} on symbiotic stars, and \cite{bhatt19} on fainter PNe. We also examined unpublished data based on M31 \ha \ targets described in \cite{AZ} (AMB). The AMB catalog of H$\alpha$ emitters often has multiple entries for large complexes, of HII regions or SNR, and the \cite{lee14} object centers are often many arcseconds from what is apparently the same object in the AMB catalog. 
Most of the non-LL14 targets are the other typical emission line sources in M31: HII regions, planetary nebulae (PNe), and symbiotic stars but many have high [S~II]/H$\alpha $ ratios and a few of those could be SNR which were not cataloged in LL14.  In this collection, there are about 2700 objects with detected [S~II]/H$\alpha $ ratios, roughly 2/3 of which are previously unpublished. Importantly, we also included a small number of these HII regions, PNe and symbiotics in our Hectochelle high resolution spectrograph project, and it is the few sources that have SNR characteristics that we report on in Section \ref{others}.

\section{Spectral reduction \label{reduction}}

These high and low resolution data were processed in the standard way used by the SAO 
Telescope Data Center (TDC\footnote{\url{https://lweb.cfa.harvard.edu/mmti/hectospec/hecto_pipe_report.pdf}}).
A global sky subtraction was done for the object spectra, using fibers typically positioned far away from the disk of M31. By not using a local background, this method meant that large scale emission could be included in the spectra of compact targets. In particular, emission lines from diffuse interstellar gas could be found in the spectra of small HII regions.

\begin{figure}[ht]
\includegraphics[width=7.5in]{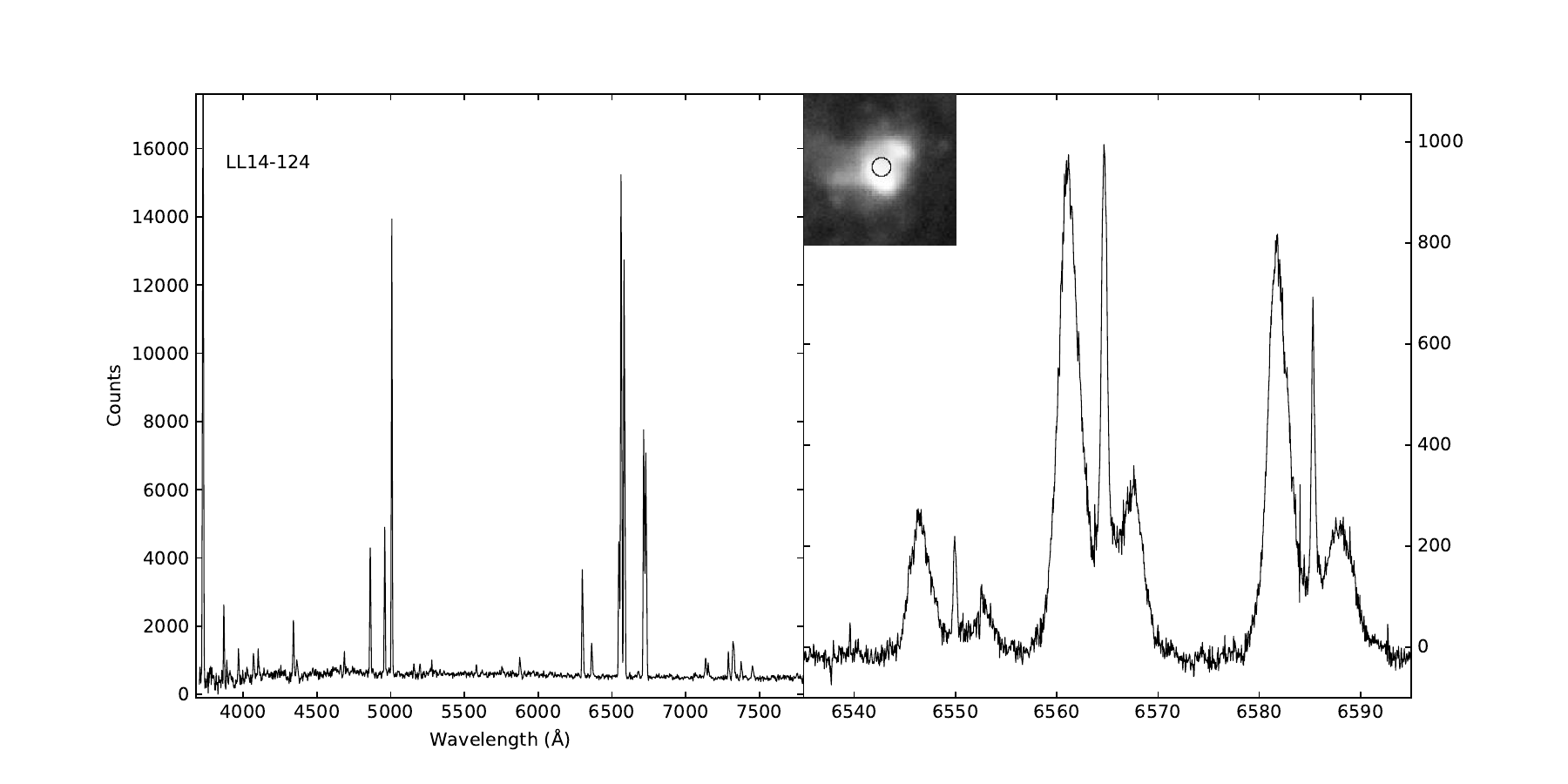} 
\caption{Example low and high resolution spectra of the young SNR LL14-124.  A postage stamp image in \ha \ is shown as an insert, from the \cite{massey} image collection. The field is 15\arcsec; a circle indicates the position and area coverage of the spectra taken with the two spectrographs. This SNR shows the complexity of some [N II]$\lambda$$\lambda$6548, 6584 and H$\alpha$ line profiles and a very large velocity width.  The [S~II]/H$\alpha$\/ ratio is 0.5, and it is detected in X-rays.  Note the narrow peaks near zero velocity (in the M31 frame) that are likely produced by a photoionization precursor. }
\label{fig:spec1}
\end{figure}

\begin{figure}[ht]
\includegraphics[width=7.5in]{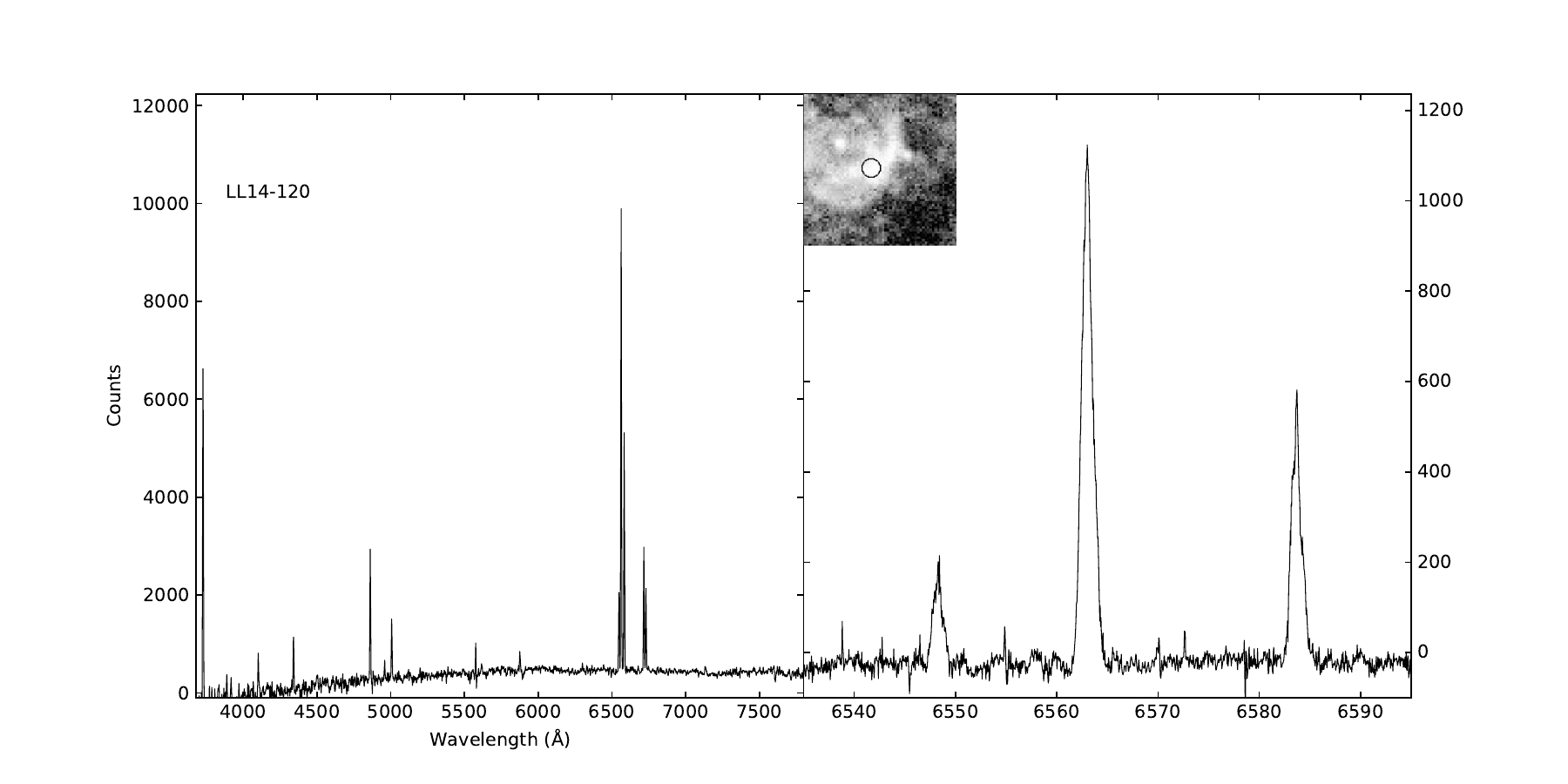} 
\caption{LL14-120, an older SNR.  With an [S~II]/H$\alpha$\/ ratio of 0.43 and line widths of 59 \kmss (FWHM) and 106 \kmss (Full Width at 20\% of peak flux).  It is confirmed as an SNR, though it is not detected in X-rays.}
\label{fig:spec2}
\end{figure}

\begin{figure}[ht]
\includegraphics[width=7.5in]{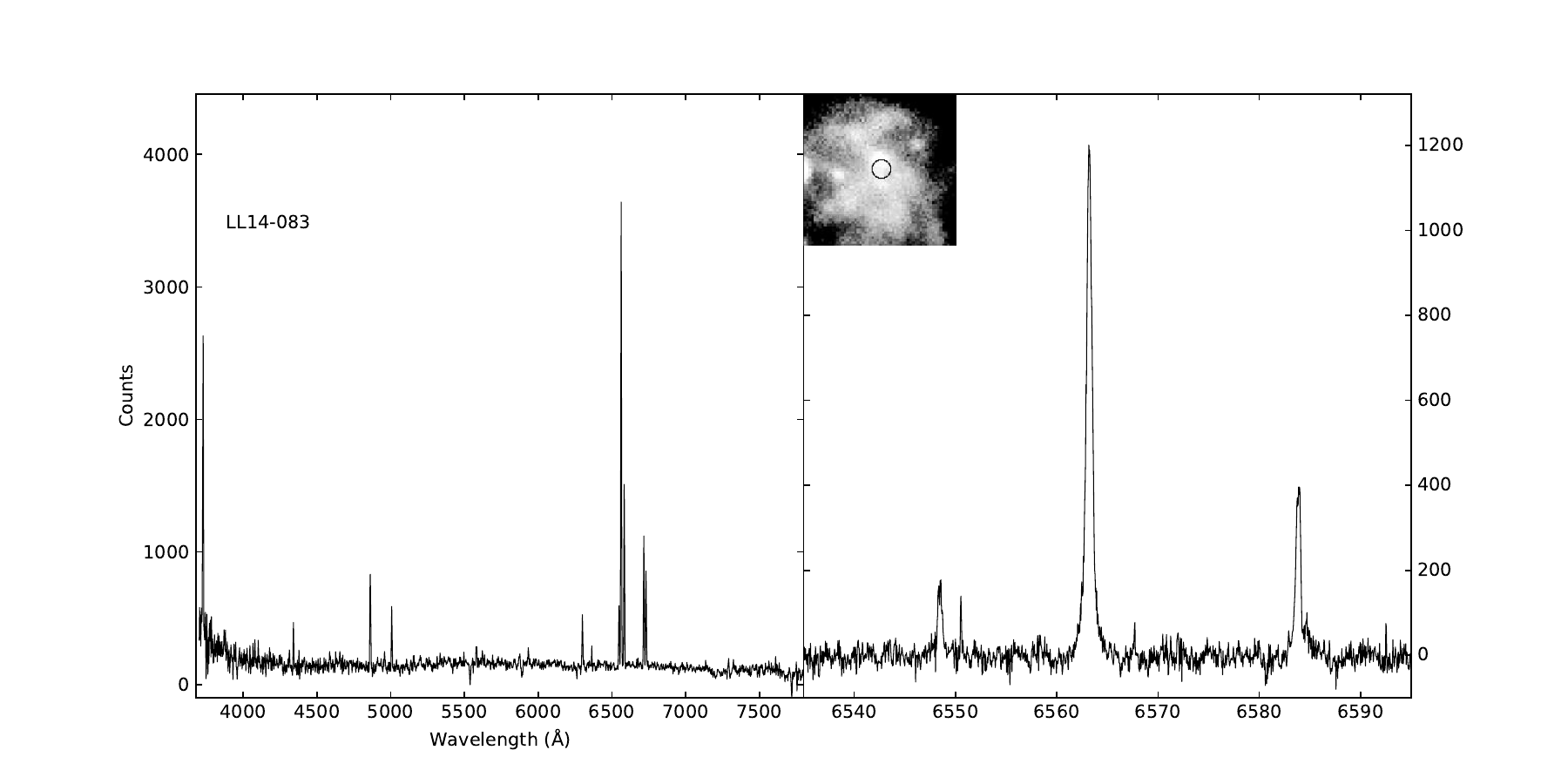} 
\caption{ LL14-83, which has narrower cores but broad wings and strong [SII]. The combination of a narrow peak with broad wings indicates a combination of shock excitation and photoionization.  The photoionized component could be a shock precursor or an H II region along the line of sight.}
\label{fig:spec3}
\end{figure}

\begin{figure}[ht]
\includegraphics[width=7.5in]{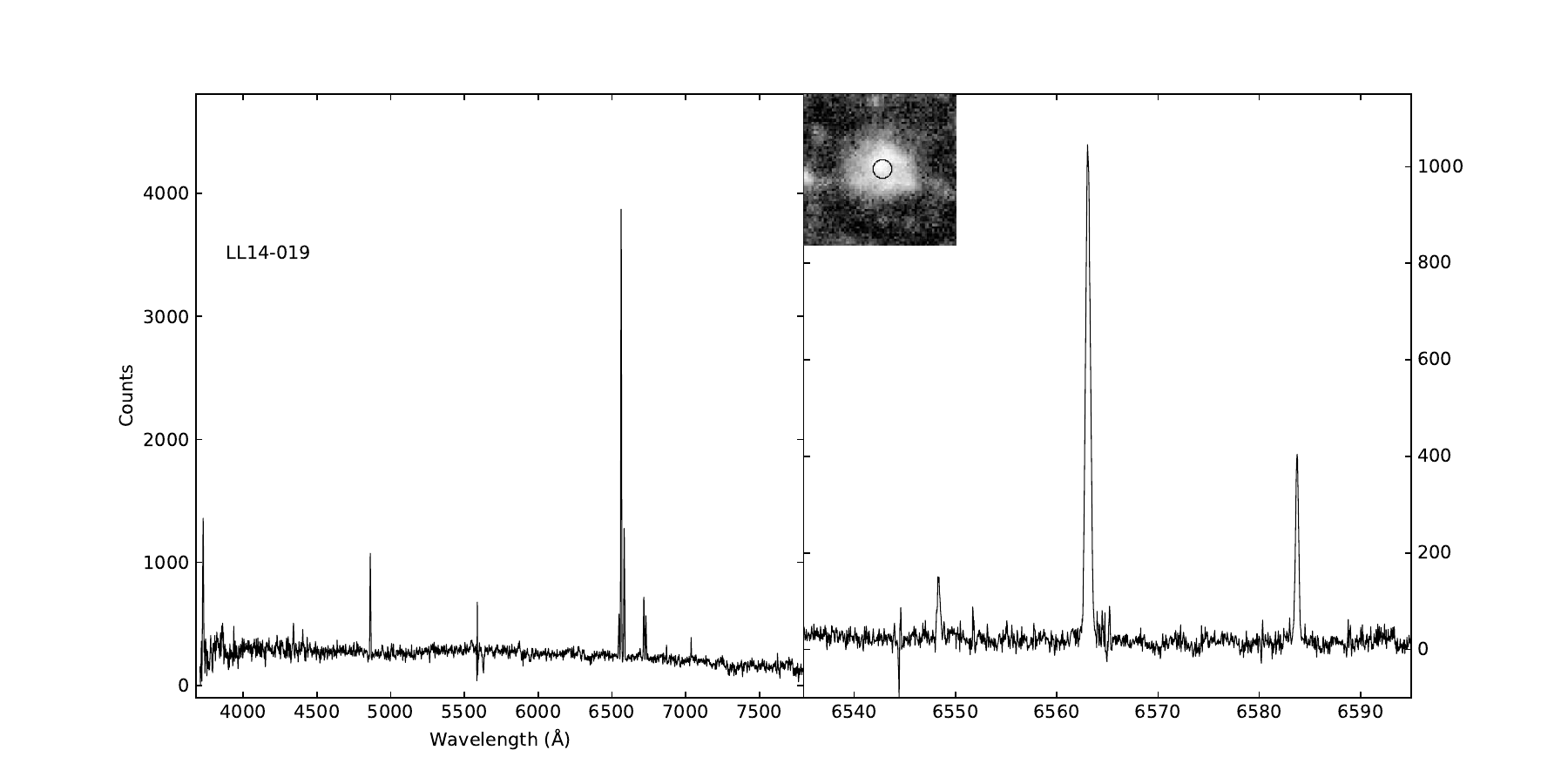} 
\caption{LL14-019, an HII region with weak [SII] and very narrow lines.  Both the narrow line width (23 \kms) and the low [S~II]/H$\alpha$\/ ratio indicate photoionized gas.}
\label{fig:spec4}
\end{figure}

All of the spectra were corrected for spectral response, but not put on a flux scale, because few of the nights were photometric. We will be working with traditional ratios of emission lines close in wavelength and with the widths of those lines, both of which are little affected by problems in flux calibration or indeed reddening, and thus we make no reddening corrections to the spectra  \citep{bpt}.  Figures~\ref{fig:spec1} to \ref{fig:spec4} show examples of high-and low-resolution spectra.  Figure~\ref{fig:spec1}, LL14-124, shows broad lines, along with
a narrow core due to photoionized gas, most likely the photoionization precursor of the SNR shock \citep{vancura92}.  Figure~\ref{fig:spec2} shows a moderately wide, single-component profile, and Figure~\ref{fig:spec3} has a narrow core with moderately
broad wings.  Figure~\ref{fig:spec4} shows a very narrow profile indicative of a photoionized nebula.


Line strengths and FWHM were measured using {\it fitprof} in IRAF for the low- and high-resolution spectra. We used a gaussian for all lines, fixed at the width determined from the H$\alpha$ line. Although clearly not appropriate for the line shapes of many of the SNRs observed at high-resolution, the gaussian values are still useful for comparisons. Errors were provided by fitprof, using a Monte-Carlo technique, but some spectra had to have continua artificially elevated to positive values. Negative continua occasionally occur with these fiber spectra, but that does not affect the line ratios as the continua are subtracted before flux calculations. The errors in those cases are affected somewhat (typically a few tens of counts needed to be added), in the sense the errors will be overestimated. Because the gaussian function does not always capture the details of many of the SNR line profiles, we elected to measure by eye the full width of the H$\alpha $\ lines at 20\% of the peak flux found in the line (FW$_{20}$).  We have no formal errors on these estimates, but believe them to be less than 10\% in the higher signal-to-noise cases, and about 30\% in the noisier cases.


The results are presented in Table \ref{ll14_spectra}.  
For each of the \citet{lee14} objects we observed, we list the line fluxes relative to H$\alpha$=300, the FWHM, and FW$_{20}$, the full width at 20\% of the peak.  Lines detected at less  than 6$\sigma$ are omitted. 
Though the fluxes measured are not comparable among objects, because the regions selected by the 1.5\arcsec\/ fibers are not representative of the entire SNRs, the uncertainties in line ratios are useful and have been measured from repeated observations. Those repeat measurements indicate rms variations in the ratios being as follows. For [NII]/\ha \ - 0.09; for [OIII]/\ha \ - 0.30; for [SII]/\ha \ - 0.13; for the much weaker [OI]/\ha \ - 0.57, where again only lines detected at greater than 6$\sigma $ were used in these statistics.

A graphical display of images, spectra and derived parameters is available in the emission-line object section of the M31 Hectospec web site \url{https://lweb.cfa.harvard.edu/oir/eg/m31clusters/M31_Hectospec.html} 
\clearpage

\begin{longrotatetable}

\begin{deluxetable}{cccccccccrr}
\tabletypesize{\scriptsize}
\decimals
\tablecaption{Hectospec spectra of SNRs (and Candidates) in M31}
\tablehead{
\colhead{Source} &
\colhead{RA} &
\colhead{Dec} &
\colhead{H$\beta$} &
\colhead{[OIII]5007} &
\colhead{[OI]6300} &
\colhead{[NII]6584} &
\colhead{[SII]6716} &
\colhead{[SII]6731} &
\colhead{FWHM} &
\colhead{FW(20)}\\
\colhead{} & \multicolumn{2}{c}{J2000}&\colhead{All fluxes scaled to H$\alpha=300$} &\colhead{}&\colhead{}&\colhead{}&\colhead{}&\colhead{}&\colhead{} \kms&\kms\\
}
\startdata
LL14-001& 9.407504& 39.86209& -- & -- & -- & -- & -- & -- & 126$\pm$ 9& 205 \\
LL14-002& 9.848688& 40.73780& 83.4$\pm$ 1.4& 47.7$\pm$ 1.5& 94.6$\pm$ 1.6& 120.6$\pm$ 1.9& 175.5$\pm$ 2.0& 125.8$\pm$ 1.7& 60$\pm$ 1& 162 \\
LL14-003& 9.878979& 40.35830& 91.2$\pm$ 4.0& 266.0 & 34.4$\pm$ 4.4& 162.3$\pm$ 5.1& 126.9$\pm$ 4.2& 81.9$\pm$ 4.4& 50$\pm$ 2& 83 \\
LL14-004& 9.937379& 40.49724& 91.4$\pm$ 4.1& 343.8 & 34.0$\pm$ 3.6& 209.6$\pm$ 4.3& 170.6$\pm$ 3.8& 114.2$\pm$ 3.4& 100$\pm$ 4& 171 \\
LL14-005& 9.961292& 40.34985& -- & -- & -- & -- & -- & -- & 29$\pm$ 1& 46 \\
LL14-006& 9.967871& 40.49709& 79.2$\pm$ 1.4& 11.1 & 12.9$\pm$ 1.3& 116.5$\pm$ 1.9& 97.6$\pm$ 1.4& 64.9$\pm$ 1.7& 32$\pm$ 1& 54 \\
LL14-007& 10.043250& 40.96058& 74.0$\pm$ 9.0& 206.2$\pm$ 9.3& 45.6$\pm$ 9.0& 168.5$\pm$ 9.3& 133.4$\pm$ 9.3& 114.7$\pm$ 9.3& 87$\pm$ 8& 165 \\
LL14-008& 10.060475& 40.62145& 21.1$\pm$ 8.4& 71.4$\pm$ 8.0& 72.4$\pm$ 8.7& 193.4$\pm$ 9.7& 191.4$\pm$ 8.0& 129.4$\pm$ 9.7& 131$\pm$ 12& 245 \\
LL14-009& 10.100463& 40.81534& 55.4$\pm$ 3.7& 7.4$\pm$ 3.5& 71.0$\pm$ 3.7& 101.0$\pm$ 4.3& 133.3$\pm$ 3.9& 94.5$\pm$ 4.5& 36$\pm$ 1& 86 \\
LL14-010& 10.126563& 40.72108& 80.6$\pm$ 3.9& 105.7$\pm$ 3.2& 26.9$\pm$ 3.1& 175.7$\pm$ 3.8& 141.7$\pm$ 3.5& 95.5$\pm$ 3.4& 51$\pm$ 1& 87 \\
LL14-011& 10.133021& 40.54647& 65.2$\pm$ 2.7& 13.0$\pm$ 2.2& 4.6$\pm$ 2.5& 82.4$\pm$ 2.6& 69.6$\pm$ 2.3& 43.5$\pm$ 2.1& 24$\pm$ 1& 40 \\
LL14-012& 10.135692& 40.50883& 60.8$\pm$ 7.2& 41.1$\pm$ 6.1& 29.4$\pm$ 5.6& 103.6$\pm$ 6.7& 143.9$\pm$ 6.1& 92.5$\pm$ 5.6& 52$\pm$ 2& 85 \\
LL14-013& 10.140942& 40.72572& 75.1$\pm$ 1.0& 8.8$\pm$ 0.7& 1.8$\pm$ 0.6& 84.6$\pm$ 0.9& 41.7$\pm$ 0.8& 29.9$\pm$ 0.7& 24$\pm$ 1& 40 \\
LL14-014& 10.140967& 40.54598& 65.1$\pm$ 0.3& 462.0$\pm$ 0.8& 5.1$\pm$ 0.2& 56.8$\pm$ 0.3& 32.9$\pm$ 0.2& 23.2$\pm$ 0.2& -- & -- \\
LL14-015& 10.149429& 40.81793& 89.6$\pm$ 1.9& 62.0$\pm$ 2.1& 12.3$\pm$ 1.9& 82.4$\pm$ 2.2& 34.3$\pm$ 1.7& 23.7$\pm$ 2.1& 26$\pm$ 1& 47 \\
LL14-016& 10.162867& 40.57927& 66.5$\pm$ 1.4& 0.9$\pm$ 1.3& 6.1$\pm$ 1.2& 127.6$\pm$ 1.8& 69.9$\pm$ 1.4& 49.7$\pm$ 1.4& 27$\pm$ 1& 44 \\
LL14-017& 10.168888& 40.81131& 33.2$\pm$ 2.5& 9.3 & 5.5$\pm$ 2.6& 87.2$\pm$ 3.1& 45.2$\pm$ 2.7& 27.7$\pm$ 2.8& 20$\pm$ 1& 35 \\
LL14-018& 10.196550& 40.92366& 53.4$\pm$ 0.5& 123.2$\pm$ 0.6& 85.4$\pm$ 0.7& 296.1$\pm$ 0.7& 226.0$\pm$ 0.8& 179.5$\pm$ 0.7& 327$\pm$ 4& 518 \\
LL14-019& 10.197196& 40.52920& 61.6$\pm$ 1.8& 8.8$\pm$ 1.8& 5.0$\pm$ 1.9& 89.8$\pm$ 2.2& 42.2$\pm$ 1.8& 27.2$\pm$ 1.6& 23$\pm$ 1& 44 \\
LL14-020& 10.262996& 41.07460& 31.1$\pm$ 2.5& 18.0$\pm$ 2.3& 102.8$\pm$ 2.8& 197.7$\pm$ 3.0& 211.6$\pm$ 2.8& 155.7$\pm$ 2.9& 91$\pm$ 4& 240 \\
LL14-021& 10.266004& 40.78447& 86.4$\pm$ 1.2& 234.4$\pm$ 1.4& 67.9$\pm$ 1.0& 267.2$\pm$ 1.6& 180.0$\pm$ 1.3& 135.6$\pm$ 1.2& 149$\pm$ 3& 231 \\
LL14-022& 10.275608& 40.86912& 91.4$\pm$ 1.1& 4.3$\pm$ 0.8& 1.0$\pm$ 0.8& 89.3$\pm$ 1.4& 36.4$\pm$ 0.9& 26.2$\pm$ 1.0& 20$\pm$ 1& 36 \\
LL14-023& 10.274229& 40.65178& 56.8$\pm$ 4.2& 301.1$\pm$ 4.3& 15.0$\pm$ 3.4& 153.6$\pm$ 4.3& 75.3$\pm$ 3.3& 52.1$\pm$ 3.3& 23$\pm$ 1& 42 \\
LL14-024& 10.295846& 40.83295& 76.9$\pm$ 1.4& 335.3$\pm$ 2.0& 24.1$\pm$ 1.3& 306.2$\pm$ 2.1& 173.4$\pm$ 1.6& 140.6$\pm$ 1.6& 36$\pm$ 1& 146 \\
LL14-025& 10.294975& 41.09993& 89.7$\pm$ 3.0& 5.4$\pm$ 3.5& 2.6$\pm$ 3.8& 101.1$\pm$ 3.5& 50.8$\pm$ 3.1& 38.6$\pm$ 2.9& 21$\pm$ 1& 31 \\
LL14-026& 10.293246& 41.12388& -- & -- & -- & -- & -- & -- & 32$\pm$ 2& 65 \\
LL14-027& 10.300779& 40.80428& -- & -- & -- & -- & -- & -- & 57$\pm$ 2& 93 \\
LL14-028& 10.300117& 41.10594& -- & -- & -- & -- & -- & -- & 25$\pm$ 1& 38 \\
LL14-029& 10.327983& 40.81846& 59.3$\pm$ 3.8& 28.3$\pm$ 4.2& 48.1$\pm$ 4.9& 182.7$\pm$ 5.1& 135.4$\pm$ 4.2& 99.4$\pm$ 4.2& 40$\pm$ 2& 77 \\
LL14-030& 10.333225& 41.29901& 107.4$\pm$ 4.5& 91.2$\pm$ 5.3& 52.3$\pm$ 4.5& 126.8$\pm$ 5.8& 109.8$\pm$ 4.5& 75.3$\pm$ 5.6& 57$\pm$ 2& 129 \\
LL14-031& 10.377417& 41.08908& 67.5$\pm$ 3.4& 264.2$\pm$ 3.5& 28.9$\pm$ 3.2& 302.4$\pm$ 4.4& 206.1$\pm$ 3.2& 145.5$\pm$ 4.1& 53$\pm$ 2& 106 \\
LL14-032& 10.394154& 41.24709& 31.2$\pm$ 12.2& 100.5 & 70.6$\pm$ 12.9& 247.1$\pm$ 16.3& 192.8$\pm$ 12.9& 116.1$\pm$ 15.6& 100$\pm$ 12& 237 \\
LL14-033& 10.398308& 41.34000& 98.3$\pm$ 4.8& 460.9$\pm$ 6.6& 15.4$\pm$ 4.3& 199.9$\pm$ 5.9& 79.4$\pm$ 4.3& 54.2$\pm$ 4.1& 23$\pm$ 1& 36 \\
LL14-034& 10.398637& 41.11550& 65.5$\pm$ 1.6& 205.5$\pm$ 2.0& 74.9$\pm$ 1.6& 360.9$\pm$ 2.4& 196.8$\pm$ 1.9& 151.4$\pm$ 1.8& 271$\pm$ 6& 424 \\
LL14-035& 10.421758& 40.63945& 69.6$\pm$ 3.5& 19.1$\pm$ 2.6& 5.9$\pm$ 2.8& 81.0$\pm$ 3.3& 62.4$\pm$ 2.6& 43.1$\pm$ 3.0& 22$\pm$ 1& 39 \\
LL14-036& 10.474796& 41.45492& 63.2$\pm$ 4.5& 9.5$\pm$ 4.2& 59.3$\pm$ 5.1& 106.4$\pm$ 5.4& 145.1$\pm$ 5.1& 100.7$\pm$ 4.8& 36$\pm$ 1& 82 \\
LL14-037& 10.544458& 40.86337& 65.5$\pm$ 5.5& 310.8$\pm$ 6.0& 46.8$\pm$ 5.3& 241.5$\pm$ 6.6& 201.9$\pm$ 4.7& 150.6$\pm$ 4.9& 130$\pm$ 5& 276 \\
LL14-038& 10.550117& 40.79184& 73.4$\pm$ 1.1& 104.1$\pm$ 1.4& 97.8$\pm$ 1.6& 203.0$\pm$ 1.7& 166.6$\pm$ 1.6& 140.6$\pm$ 1.3& 113$\pm$ 4& 187 \\
LL14-039& 10.561042& 40.86813& 43.3$\pm$ 1.4& 124.1$\pm$ 1.9& 78.2$\pm$ 2.1& 227.0$\pm$ 2.3& 204.9$\pm$ 2.2& 156.0$\pm$ 1.8& -- & -- \\
LL14-040& 10.583417& 41.46622& 56.3$\pm$ 0.9& 23.9$\pm$ 0.9& 13.1$\pm$ 0.9& 129.7$\pm$ 1.5& 80.5$\pm$ 1.1& 59.3$\pm$ 1.0& 47$\pm$ 1& 81 \\
LL14-041& 10.602813& 41.29155& 41.5$\pm$ 6.9& 373.2$\pm$ 10.2& 68.5$\pm$ 9.5& 426.7$\pm$ 11.2& 209.8$\pm$ 8.6& 161.4$\pm$ 10.8& 133$\pm$ 3& 230 \\
LL14-042& 10.605837& 40.87469& 60.9$\pm$ 7.2& 164.2$\pm$ 9.2& 51.7$\pm$ 8.0& 185.0$\pm$ 9.6& 157.8$\pm$ 8.0& 121.0$\pm$ 9.2& 103$\pm$ 9& 184 \\
LL14-043& 10.607042& 41.29691& -- & -- & -- & -- & -- & -- & 130$\pm$ 3& 186 \\
LL14-044& 10.612646& 41.49233& 51.0$\pm$ 3.2& 9.7$\pm$ 2.8& 10.2$\pm$ 3.4& 113.3$\pm$ 4.4& 78.5$\pm$ 3.8& 58.6$\pm$ 3.4& 49$\pm$ 2& 71 \\
LL14-045& 10.630992& 41.10165& 64.9$\pm$ 2.4& 30.9$\pm$ 2.7& 42.4$\pm$ 3.0& 184.9$\pm$ 3.8& 152.9$\pm$ 2.7& 105.8$\pm$ 3.4& 135$\pm$ 2& 211 \\
LL14-046& 10.685454& 40.91051& 74.7$\pm$ 2.2& 267.5$\pm$ 2.5& 33.1$\pm$ 1.8& 148.8$\pm$ 2.5& 109.9$\pm$ 1.9& 78.4$\pm$ 1.8& 68$\pm$ 3& 145 \\
LL14-047& 10.698654& 41.02315& 62.4$\pm$ 4.4& 142.4$\pm$ 3.8& 73.3$\pm$ 4.0& 223.1$\pm$ 5.3& 186.6$\pm$ 3.8& 129.4$\pm$ 4.9& 58$\pm$ 1& 137 \\
LL14-048& 10.704167& 41.40108& 71.2$\pm$ 3.4& 495.8$\pm$ 5.3& 25.5$\pm$ 4.0& 338.0$\pm$ 5.6& 220.9$\pm$ 4.2& 174.3$\pm$ 3.8& 161$\pm$ 3& 176 \\
LL14-049& 10.704071& 41.56530& 51.9$\pm$ 1.1& 9.0$\pm$ 0.9& 8.7$\pm$ 0.9& 88.2$\pm$ 1.6& 66.1$\pm$ 1.4& 45.4$\pm$ 1.2& 46$\pm$ 1& 124 \\
LL14-050& 10.723475& 41.43090& 96.2$\pm$ 0.4& 163.5$\pm$ 0.5& 96.8$\pm$ 0.5& 336.4$\pm$ 0.6& 184.5$\pm$ 0.5& 178.1$\pm$ 0.5& 201$\pm$ 1& 384 \\
LL14-051& 10.730612& 40.99570& 64.3$\pm$ 3.5& 342.3$\pm$ 6.0& 30.6$\pm$ 4.7& 221.1$\pm$ 4.8& 151.4$\pm$ 4.7& 130.0$\pm$ 4.3& 28$\pm$ 1& 110 \\
LL14-052& 10.732788& 41.43837& 77.7$\pm$ 6.1& 148.9$\pm$ 6.1& 83.1$\pm$ 6.1& 336.3$\pm$ 7.3& 172.4$\pm$ 5.6& 128.6$\pm$ 5.4& 97$\pm$ 3& 175 \\
LL14-053& 10.737200& 41.58506& 66.4$\pm$ 4.7& 22.0$\pm$ 4.5& 23.8$\pm$ 4.5& 148.4$\pm$ 5.5& 116.4$\pm$ 5.2& 80.3$\pm$ 4.5& 31$\pm$ 1& 100 \\
LL14-054& 10.766229& 41.60626& 94.1$\pm$ 1.6& 50.7$\pm$ 1.7& 23.8$\pm$ 1.4& 142.4$\pm$ 2.4& 124.1$\pm$ 1.9& 89.8$\pm$ 1.7& 49$\pm$ 1& 92 \\
LL14-055& 10.768329& 41.08157& 45.9$\pm$ 2.3& 80.1$\pm$ 2.0& 7.4$\pm$ 1.6& 97.3$\pm$ 2.3& 69.2$\pm$ 2.1& 48.1$\pm$ 1.8& 43$\pm$ 1& 80 \\
LL14-056& 10.776917& 41.64761& 66.0$\pm$ 1.9& 15.3$\pm$ 1.8& 4.0$\pm$ 1.8& 123.4$\pm$ 2.8& 73.6$\pm$ 2.0& 52.8$\pm$ 2.0& 29$\pm$ 1& 57 \\
LL14-057& 10.786729& 41.10056& 37.9$\pm$ 3.4& 19.9$\pm$ 2.7& 15.5$\pm$ 3.4& 105.6$\pm$ 3.7& 93.1$\pm$ 3.4& 62.2$\pm$ 3.0& 49$\pm$ 1& 88 \\
LL14-058& 10.786583& 41.08609& 59.1$\pm$ 6.0& 191.1 & 60.8$\pm$ 5.8& 232.1$\pm$ 6.3& 175.3$\pm$ 5.6& 123.4$\pm$ 5.0& 54$\pm$ 2& 90 \\
LL14-059& 10.785625& 41.05280& 52.5$\pm$ 6.8& 372.2$\pm$ 10.4& 37.3$\pm$ 8.0& 273.1$\pm$ 10.8& 155.2$\pm$ 8.0& 118.7$\pm$ 10.4& 55$\pm$ 6& 126 \\
LL14-060& 10.795029& 41.62821& 49.4$\pm$ 2.5& 2.2$\pm$ 1.7& 8.5$\pm$ 1.7& 124.8$\pm$ 2.7& 68.8$\pm$ 1.8& 50.0$\pm$ 2.4& 34$\pm$ 1& 64 \\
LL14-061& 10.797512& 41.48499& 75.3$\pm$ 4.1& 233.3$\pm$ 4.1& 64.6$\pm$ 3.7& 326.7$\pm$ 4.0& 192.7$\pm$ 3.4& 137.6$\pm$ 3.4& 102$\pm$ 3& 174 \\
LL14-062& 10.810158& 40.90744& 77.1$\pm$ 0.8& 4.4$\pm$ 0.7& 2.6$\pm$ 0.6& 76.4$\pm$ 1.0& 45.4$\pm$ 0.7& 30.6$\pm$ 0.9& 24$\pm$ 1& 38 \\
LL14-063& 10.832558& 41.05155& 63.2$\pm$ 3.3& 17.4$\pm$ 2.4& 6.1$\pm$ 2.7& 94.2$\pm$ 3.3& 35.5$\pm$ 2.6& 30.0$\pm$ 2.4& 26$\pm$ 1& 49 \\
LL14-064& 10.846388& 41.11156& 62.5$\pm$ 11.4& 382.5$\pm$ 10.0& 60.6$\pm$ 11.4& 279.5$\pm$ 12.4& 175.0$\pm$ 9.5& 123.5$\pm$ 9.5& 107$\pm$ 11& 191 \\
LL14-065& 10.845575& 41.09743& -- & -- & -- & -- & -- & -- & 38$\pm$ 1& 75 \\
LL14-066& 10.865879& 41.30923& 85.4$\pm$ 0.9& 163.9$\pm$ 1.0& 76.6$\pm$ 1.1& 342.5$\pm$ 1.3& 171.7$\pm$ 1.0& 145.0$\pm$ 0.9& 167$\pm$ 1& 285 \\
LL14-067& 10.873650& 41.31850& -- & -- & -- & -- & -- & -- & 42$\pm$ 1& 76 \\
LL14-068& 10.899979& 41.23546& -- & -- & -- & -- & -- & -- & 49$\pm$ 4& 80 \\
LL14-069& 10.908100& 41.73786& 79.0$\pm$ 3.6& -0.5 & 7.9$\pm$ 2.6& 88.1$\pm$ 3.0& 54.0$\pm$ 2.7& 37.0$\pm$ 2.4& 24$\pm$ 1& 41 \\
LL14-070& 10.913333& 41.44825& 99.3$\pm$ 0.6& 319.8$\pm$ 0.8& 83.1$\pm$ 0.7& 383.4$\pm$ 0.8& 188.3$\pm$ 0.6& 178.3$\pm$ 0.6& 208$\pm$ 1& 412 \\
LL14-071& 10.919400& 41.18200& 55.3$\pm$ 3.4& 118.8$\pm$ 4.0& 66.8$\pm$ 4.3& 202.1$\pm$ 5.1& 189.9$\pm$ 3.8& 129.2$\pm$ 3.8& 90$\pm$ 2& 181 \\
LL14-072& 10.924933& 41.76389& 62.1$\pm$ 2.6& 14.6 & 1.5$\pm$ 2.2& 87.4$\pm$ 3.1& 37.8$\pm$ 2.5& 26.7$\pm$ 2.2& 22$\pm$ 1& 27 \\
LL14-073& 10.939879& 41.44716& $-$34.4$\pm$ 30.6& 150.5$\pm$ 37.1& 108.6$\pm$ 39.2& 317.7$\pm$ 39.2& 203.8$\pm$ 32.3& 184.4$\pm$ 42.5& 45$\pm$ 1& 126 \\
LL14-074& 10.941554& 41.74226& 66.8$\pm$ 2.8& 6.2$\pm$ 2.9& 9.1$\pm$ 2.6& 87.9$\pm$ 3.4& 42.7$\pm$ 2.8& 34.6$\pm$ 2.6& 23$\pm$ 1& 44 \\
LL14-075& 10.947575& 41.21558& 68.9$\pm$ 12.1& 76.3$\pm$ 9.8& 56.8$\pm$ 10.1& 189.3$\pm$ 13.5& 125.4$\pm$ 9.8& 80.0$\pm$ 13.5& 47$\pm$ 2& 80 \\
LL14-076& 10.973337& 41.19990& 81.3$\pm$ 1.7& 283.0$\pm$ 2.6& 31.1$\pm$ 1.8& 204.3$\pm$ 2.4& 140.3$\pm$ 2.1& 104.7$\pm$ 1.8& 96$\pm$ 10& 194 \\
LL14-077& 10.975783& 41.68909& 69.8$\pm$ 2.8& 14.7$\pm$ 2.8& 4.6$\pm$ 2.6& 141.4$\pm$ 4.0& 58.1$\pm$ 3.1& 42.2$\pm$ 2.6& 22$\pm$ 1& 37 \\
LL14-078& 10.977471& 41.88104& 85.7$\pm$ 1.3& 241.2$\pm$ 2.0& 53.5$\pm$ 1.2& 192.4$\pm$ 1.8& 157.7$\pm$ 1.7& 120.2$\pm$ 1.5& 134$\pm$ 1& 341 \\
LL14-079& 10.983592& 41.78823& 67.5$\pm$ 2.6& 82.0$\pm$ 2.9& 24.3$\pm$ 2.6& 199.0$\pm$ 4.2& 110.4$\pm$ 2.8& 80.2$\pm$ 3.5& 44$\pm$ 1& 77 \\
LL14-080& 10.992812& 41.22486& 82.6$\pm$ 1.2& 202.7$\pm$ 1.4& 56.7$\pm$ 1.3& 235.0$\pm$ 1.5& 201.0$\pm$ 1.7& 147.1$\pm$ 1.4& 288$\pm$ 5& 351 \\
LL14-081& 10.998342& 41.80768& 66.7$\pm$ 3.3& 113.1$\pm$ 3.6& 16.3$\pm$ 2.9& 117.5$\pm$ 4.0& 56.4$\pm$ 3.3& 37.2$\pm$ 2.9& 27$\pm$ 1& 53 \\
LL14-082& 11.004537& 41.35158& 67.7$\pm$ 2.6& 543.6$\pm$ 3.1& 40.9$\pm$ 2.9& 346.7$\pm$ 3.5& 143.0$\pm$ 2.6& 126.3$\pm$ 2.2& 139$\pm$ 3& 318 \\
LL14-083& 11.019050& 41.81416& 61.9$\pm$ 1.7& 41.3$\pm$ 1.8& 31.0$\pm$ 1.8& 118.6$\pm$ 2.6& 83.0$\pm$ 1.9& 61.3$\pm$ 1.8& 32$\pm$ 1& 66 \\
LL14-084& 11.021221& 41.45471& 53.2$\pm$ 2.1& 4.3$\pm$ 1.4& 1.2$\pm$ 1.7& 103.0$\pm$ 2.1& 62.2$\pm$ 1.7& 45.0$\pm$ 1.6& 38$\pm$ 1& 79 \\
LL14-085& 11.022338& 41.33486& -- & -- & -- & -- & -- & -- & 53$\pm$ 1& 102 \\
LL14-086& 11.053417& 41.84238& 61.8$\pm$ 2.6& 2.2$\pm$ 2.8& $-$4.1$\pm$ 2.1& 58.9$\pm$ 3.2& 38.6$\pm$ 2.5& 27.6$\pm$ 3.2& 19$\pm$ 1& 41 \\
LL14-087& 11.055650& 41.33174& 52.7$\pm$ 0.7& 77.9$\pm$ 0.7& 96.6$\pm$ 0.9& 270.6$\pm$ 1.2& 190.2$\pm$ 0.8& 178.0$\pm$ 1.0& -- & -- \\
LL14-088& 11.071675& 41.86124& 68.0$\pm$ 7.4& 65.9$\pm$ 7.4& 63.8$\pm$ 7.4& 251.0$\pm$ 7.9& 195.1$\pm$ 7.9& 140.2$\pm$ 7.4& 107$\pm$ 5& 189 \\
LL14-089& 11.085746& 41.30171& 71.8$\pm$ 1.5& 71.0$\pm$ 1.7& 2.1$\pm$ 1.4& 92.0$\pm$ 1.8& 57.1$\pm$ 1.6& 42.0$\pm$ 1.3& 29$\pm$ 1& 68 \\
LL14-090& 11.084400& 41.58070& 82.8$\pm$ 3.7& 2.7$\pm$ 4.5& 4.1$\pm$ 3.3& 100.5$\pm$ 4.5& 58.1$\pm$ 3.7& 39.3$\pm$ 3.5& 23$\pm$ 1& 37 \\
LL14-091& 11.095546& 41.30803& -- & -- & -- & -- & -- & -- & 32$\pm$ 1& 57 \\
LL14-092& 11.108379& 41.41437& 30.3$\pm$ 2.4& 10.2$\pm$ 2.7& 7.1$\pm$ 2.5& 112.6$\pm$ 3.4& 62.6$\pm$ 2.7& 45.5$\pm$ 2.7& 27$\pm$ 1& 47 \\
LL14-093& 11.110321& 41.81567& 61.2$\pm$ 6.4& 15.0$\pm$ 5.4& 23.0$\pm$ 5.1& 114.8$\pm$ 6.1& 85.4$\pm$ 5.4& 64.1$\pm$ 4.8& -- & -- \\
LL14-094& 11.117375& 41.30282& 73.8$\pm$ 3.7& 60.5$\pm$ 4.4& 76.2$\pm$ 4.1& 194.1$\pm$ 5.5& 158.3$\pm$ 4.1& 110.9$\pm$ 4.1& 63$\pm$ 1& 128 \\
LL14-095& 11.122867& 41.87889& 66.3$\pm$ 1.5& 19.2$\pm$ 1.4& 5.5$\pm$ 1.4& 105.6$\pm$ 2.0& 43.8$\pm$ 1.4& 29.7$\pm$ 1.4& 29$\pm$ 1& 51 \\
LL14-096& 11.134133& 41.39407& 82.2$\pm$ 1.4& 7.4$\pm$ 0.7& 7.0$\pm$ 0.8& 108.5$\pm$ 1.4& 60.4$\pm$ 0.9& 42.7$\pm$ 1.1& 37$\pm$ 1& 65 \\
LL14-097& 11.149408& 41.42369& 80.5$\pm$ 1.7& 37.9$\pm$ 1.3& 60.1$\pm$ 1.5& 179.4$\pm$ 1.9& 161.7$\pm$ 1.5& 112.8$\pm$ 1.4& 52$\pm$ 1& 107 \\
LL14-098& 11.152554& 41.41656& 83.5$\pm$ 2.6& 32.4$\pm$ 2.5& 25.0$\pm$ 2.9& 172.3$\pm$ 3.2& 146.8$\pm$ 2.6& 99.6$\pm$ 2.9& 96$\pm$ 1& 175 \\
LL14-100& 11.161775& 41.42516& -- & -- & -- & -- & -- & -- & 57$\pm$ 1& 123 \\
LL14-101& 11.172904& 41.46552& 55.2$\pm$ 2.6& 19.7$\pm$ 2.8& 22.3$\pm$ 2.9& 149.5$\pm$ 3.9& 131.9$\pm$ 3.6& 95.8$\pm$ 2.9& 63$\pm$ 1& 102 \\
LL14-102& 11.181088& 41.44817& 89.0$\pm$ 3.5& 47.8$\pm$ 2.7& 19.1$\pm$ 2.7& 114.9$\pm$ 3.2& 96.9$\pm$ 2.7& 64.6$\pm$ 2.4& 52$\pm$ 1& 90 \\
LL14-103& 11.180100& 41.43763& 93.0$\pm$ 1.7& 97.0$\pm$ 1.6& 35.3$\pm$ 1.4& 135.0$\pm$ 2.1& 119.1$\pm$ 1.5& 84.8$\pm$ 1.4& 60$\pm$ 1& 127 \\
LL14-104& 11.183858& 41.96564& 72.9$\pm$ 2.7& 46.4$\pm$ 2.6& 67.9$\pm$ 2.7& 123.9$\pm$ 3.2& 138.5$\pm$ 2.9& 98.4$\pm$ 2.6& 45$\pm$ 1& 105 \\
LL14-105& 11.191725& 41.88311& 80.2$\pm$ 2.6& 463.2$\pm$ 3.7& 57.9$\pm$ 2.5& 307.3$\pm$ 3.8& 169.0$\pm$ 2.7& 140.7$\pm$ 2.6& 125$\pm$ 3& 192 \\
LL14-106& 11.196192& 41.48905& 84.7$\pm$ 2.1& 75.4$\pm$ 2.2& 67.2$\pm$ 2.5& 187.2$\pm$ 2.9& 168.3$\pm$ 2.6& 120.7$\pm$ 2.1& 80$\pm$ 1& 164 \\
LL14-107& 11.208108& 41.88413& -- & -- & -- & -- & -- & -- & 31$\pm$ 1& 116 \\
LL14-108& 11.209700& 41.46535& 59.1$\pm$ 4.8& 119.6$\pm$ 4.8& 70.4$\pm$ 5.2& 222.1$\pm$ 6.5& 152.7$\pm$ 4.2& 114.4$\pm$ 5.7& 37$\pm$ 2& 209 \\
LL14-109& 11.211254& 41.90506& 75.0$\pm$ 4.3& 136.9$\pm$ 5.7& 78.7$\pm$ 5.4& 224.1$\pm$ 6.6& 136.9$\pm$ 4.8& 107.8$\pm$ 6.3& 83$\pm$ 2& 169 \\
LL14-110& 11.209900& 41.53576& 59.0$\pm$ 0.6& 26.3$\pm$ 0.5& 1.4$\pm$ 0.6& 87.9$\pm$ 0.8& 27.6$\pm$ 0.5& 19.2$\pm$ 0.5& 25$\pm$ 1& 44 \\
LL14-111& 11.212692& 41.48424& 92.0$\pm$ 2.4& 229.4$\pm$ 3.1& 15.4$\pm$ 1.8& 197.6$\pm$ 3.3& 148.5$\pm$ 2.6& 117.2$\pm$ 2.1& 117$\pm$ 2& 232 \\
LL14-112& 11.219958& 41.91655& 50.0$\pm$ 1.0& 39.2$\pm$ 1.0& 93.9$\pm$ 1.1& 234.3$\pm$ 1.4& 214.8$\pm$ 1.3& 156.9$\pm$ 1.1& 140$\pm$ 1& 252 \\
LL14-113& 11.225600& 41.53083& 66.6$\pm$ 3.5& 18.9$\pm$ 3.9& 44.2$\pm$ 4.2& 103.6$\pm$ 5.0& 108.6$\pm$ 4.4& 75.4$\pm$ 3.9& 75$\pm$ 2& 129 \\
LL14-114& 11.232058& 41.94995& 56.6$\pm$ 7.2& 261.1$\pm$ 7.7& 61.1$\pm$ 7.2& 218.1$\pm$ 9.0& 128.1$\pm$ 7.7& 105.9$\pm$ 6.8& 63$\pm$ 4& 129 \\
LL14-115& 11.250771& 41.47409& 55.2$\pm$ 4.6& 26.9$\pm$ 4.4& 26.6$\pm$ 4.2& 123.5$\pm$ 4.6& 115.3$\pm$ 3.8& 89.2$\pm$ 4.8& 61$\pm$ 1& 116 \\
LL14-116& 11.256317& 41.99142& 69.9$\pm$ 3.0& 8.0 & 11.1$\pm$ 2.9& 74.8$\pm$ 3.2& 36.4$\pm$ 2.5& 27.4$\pm$ 2.5& 22$\pm$ 1& 32 \\
LL14-117& 11.271937& 41.64872& 64.2$\pm$ 3.2& 9.5$\pm$ 2.3& 8.2$\pm$ 2.6& 86.0$\pm$ 2.9& 51.2$\pm$ 2.3& 39.0$\pm$ 2.4& 19$\pm$ 1& 33 \\
LL14-118& 11.282592& 41.59662& 79.4$\pm$ 7.1& 62.6$\pm$ 6.0& 34.7$\pm$ 5.2& 178.9$\pm$ 7.1& 159.9$\pm$ 6.0& 107.3$\pm$ 6.7& 41$\pm$ 1& 77 \\
LL14-119& 11.281467& 41.54013& 46.8$\pm$ 1.1& 26.6$\pm$ 1.1& 5.7$\pm$ 1.2& 102.7$\pm$ 1.4& 41.5$\pm$ 1.1& 28.2$\pm$ 1.2& 26$\pm$ 1& 47 \\
LL14-120& 11.288854& 41.85220& 83.6$\pm$ 1.2& 37.7$\pm$ 1.1& 4.2$\pm$ 1.2& 152.5$\pm$ 1.6& 79.2$\pm$ 1.0& 52.7$\pm$ 1.4& 59$\pm$ 1& 106 \\
LL14-121& 11.297292& 41.66886& -- & -- & -- & -- & -- & -- & 22$\pm$ 1& 40 \\
LL14-122& 11.299254& 41.83637& 51.3$\pm$ 5.2& $-$6.3$\pm$ 5.2& $-$5.2$\pm$ 5.4& 101.0$\pm$ 5.6& 44.1$\pm$ 4.5& 34.7$\pm$ 5.4& 22$\pm$ 1& 45 \\
LL14-123& 11.306971& 41.59592& 40.4$\pm$ 4.9& 87.7$\pm$ 6.2& 71.7$\pm$ 6.5& 172.2$\pm$ 6.8& 176.1$\pm$ 6.5& 135.0$\pm$ 5.5& 73$\pm$ 3& 150 \\
LL14-124& 11.308192& 41.60436& 65.2$\pm$ 0.5& 191.2$\pm$ 0.7& 57.3$\pm$ 0.6& 238.8$\pm$ 0.7& 147.7$\pm$ 0.7& 144.0$\pm$ 0.6& 353$\pm$ 3& 467 \\
LL14-125& 11.315242& 41.57154& 59.0$\pm$ 5.7& 11.7$\pm$ 7.2& 29.5$\pm$ 6.0& 119.2$\pm$ 7.2& 107.4$\pm$ 6.0& 62.0$\pm$ 7.2& 29$\pm$ 1& 50 \\
LL14-126& 11.321446& 41.79090& 31.1$\pm$ 6.8& 15.1$\pm$ 8.8& $-$7.8$\pm$ 9.2& 119.1$\pm$ 9.7& 48.6$\pm$ 8.8& 45.7$\pm$ 9.7& 21$\pm$ 1& 37 \\
LL14-127& 11.328233& 41.90710& 64.3$\pm$ 5.2& 11.2$\pm$ 3.4& 22.2$\pm$ 4.4& 145.2$\pm$ 5.2& 58.8$\pm$ 3.9& 41.6$\pm$ 4.4& 24$\pm$ 1& 44 \\
LL14-128& 11.340433& 41.66676& 80.7$\pm$ 3.8& 173.3$\pm$ 5.1& 43.0$\pm$ 4.3& 241.1$\pm$ 6.2& 182.4$\pm$ 4.6& 125.7$\pm$ 5.4& 36$\pm$ 1& 62 \\
LL14-129& 11.353171& 41.72716& 53.6$\pm$ 2.4& 45.2$\pm$ 2.9& 81.1$\pm$ 3.5& 185.1$\pm$ 4.3& 150.0$\pm$ 3.6& 108.0$\pm$ 2.9& 113$\pm$ 2& 198 \\
LL14-130& 11.357358& 41.68294& -- & -- & -- & -- & -- & -- & 21$\pm$ 1& 34 \\
LL14-131& 11.357379& 41.72279& -- & -- & -- & -- & -- & -- & 22$\pm$ 1& 39 \\
LL14-132& 11.360650& 41.71447& 75.2$\pm$ 4.1& 155.8$\pm$ 5.4& 12.6$\pm$ 4.9& 87.8$\pm$ 5.6& 39.3$\pm$ 4.1& 36.7$\pm$ 5.1& 22$\pm$ 1& 36 \\
LL14-133& 11.364958& 41.71810& 43.2$\pm$ 4.7& 121.5$\pm$ 5.7& 65.4$\pm$ 6.0& 227.6$\pm$ 6.7& 155.7$\pm$ 5.5& 105.6$\pm$ 6.7& 85$\pm$ 3& 167 \\
LL14-134& 11.364363& 41.87238& 48.4$\pm$ 6.2& 103.4 & 0.6$\pm$ 8.5& 58.9$\pm$ 5.9& 27.5$\pm$ 5.1& 17.6$\pm$ 6.5& 41$\pm$ 1& 57 \\
LL14-135& 11.364387& 41.77537& 62.0$\pm$ 1.8& 16.0 & 3.3$\pm$ 1.5& 106.0$\pm$ 2.5& 59.9$\pm$ 1.7& 42.8$\pm$ 2.2& 32$\pm$ 1& 47 \\
LL14-136& 11.371237& 41.77431& 4.4$\pm$ 7.6& 16.8$\pm$ 6.1& 15.9$\pm$ 6.1& 115.6$\pm$ 6.3& 79.9$\pm$ 5.7& 62.5$\pm$ 5.9& 39$\pm$ 1& 64 \\
LL14-137& 11.374063& 41.78952& 33.3$\pm$ 6.5& 32.1$\pm$ 7.7& 73.1$\pm$ 9.2& 145.4$\pm$ 9.2& 169.9$\pm$ 8.4& 121.7$\pm$ 7.7& 45$\pm$ 1& 105 \\
LL14-138& 11.384483& 41.80273& 74.4$\pm$ 3.4& 10.5$\pm$ 3.8& 7.5$\pm$ 3.7& 117.8$\pm$ 4.1& 66.2$\pm$ 3.6& 51.1$\pm$ 3.5& 22$\pm$ 1& 44 \\
LL14-139& 11.398017& 41.96806& 56.4$\pm$ 2.5& 232.3$\pm$ 2.7& 26.5$\pm$ 2.2& 189.2$\pm$ 2.7& 130.0$\pm$ 2.3& 103.9$\pm$ 2.1& 71$\pm$ 1& 163 \\
LL14-140& 11.403596& 41.89996& 57.3$\pm$ 3.2& 52.4$\pm$ 2.7& 23.6$\pm$ 2.8& 143.4$\pm$ 3.3& 101.0$\pm$ 2.7& 71.9$\pm$ 2.6& 51$\pm$ 1& 98 \\
LL14-141& 11.403004& 41.79823& 47.3$\pm$ 1.7& 14.1 & 6.4$\pm$ 1.9& 91.7$\pm$ 2.2& 56.2$\pm$ 1.9& 39.3$\pm$ 1.8& 26$\pm$ 1& 42 \\
LL14-142& 11.409529& 41.84032& 77.1$\pm$ 2.8& 1.6$\pm$ 2.7& 13.3$\pm$ 2.8& 91.4$\pm$ 3.0& 79.2$\pm$ 2.5& 52.4$\pm$ 2.9& 29$\pm$ 1& 51 \\
LL14-143& 11.412475& 41.78679& 26.9$\pm$ 4.7& 32.6 & 29.6$\pm$ 5.3& 137.9$\pm$ 6.6& 155.2$\pm$ 6.3& 113.6$\pm$ 5.6& 50$\pm$ 1& 116 \\
LL14-144& 11.430846& 41.93341& 82.9$\pm$ 2.0& 18.0$\pm$ 1.4& 15.2$\pm$ 1.6& 136.6$\pm$ 2.0& 118.4$\pm$ 1.6& 88.3$\pm$ 1.8& 30$\pm$ 1& 54 \\
LL14-145& 11.485058& 42.18395& 99.4$\pm$ 1.2& 33.0 & 9.6$\pm$ 0.9& 117.3$\pm$ 1.5& 68.6$\pm$ 1.3& 46.9$\pm$ 1.0& 40$\pm$ 1& 57 \\
LL14-146& 11.517383& 41.83308& 75.7$\pm$ 4.4& 6.4 & 14.7$\pm$ 5.3& 92.3$\pm$ 5.5& 70.4$\pm$ 4.7& 49.6$\pm$ 6.1& 23$\pm$ 1& 37 \\
LL14-147& 11.583867& 41.88543& 75.5$\pm$ 4.2& 39.4 & 10.3$\pm$ 4.0& 83.8$\pm$ 4.2& 35.2$\pm$ 4.0& 23.3$\pm$ 4.0& 24$\pm$ 1& 32 \\
LL14-148& 11.623967& 41.97075& 73.0$\pm$ 2.2& 5.9 & 27.3$\pm$ 2.5& 104.6$\pm$ 2.9& 98.6$\pm$ 2.5& 70.2$\pm$ 2.2& 34$\pm$ 1& 52 \\
LL14-149& 11.634175& 41.99657& 70.9$\pm$ 2.1& 14.5 & 5.4$\pm$ 1.6& 124.3$\pm$ 2.2& 55.3$\pm$ 1.6& 39.5$\pm$ 1.9& 22$\pm$ 1& 32 \\
LL14-150& 11.642504& 42.17992& 94.4$\pm$ 1.6& 8.7 & 4.8$\pm$ 1.1& 106.7$\pm$ 1.9& 73.4$\pm$ 1.5& 54.0$\pm$ 1.4& 38$\pm$ 1& 57 \\
LL14-152& 11.648071& 42.22585& 94.0$\pm$ 2.3& 21.6$\pm$ 1.6& 22.8$\pm$ 1.6& 122.4$\pm$ 2.3& 89.2$\pm$ 1.6& 61.9$\pm$ 2.1& -- & -- \\
LL14-154& 11.662562& 42.12250& -- & -- & -- & -- & -- & -- & 129$\pm$ 10& 232 \\
LL14-155& 11.666100& 42.18972& 82.8$\pm$ 3.6& 82.5 & 33.7$\pm$ 3.3& 164.1$\pm$ 4.8& 136.1$\pm$ 4.1& 91.4$\pm$ 3.8& 64$\pm$ 2& 94 \\
\enddata
\label{ll14_spectra}
\end{deluxetable}

\end{longrotatetable}

\clearpage



\section{Identifying SNRs \label{criteria}}

Supernova remnants are nebulae that are observed optically as the result of shocks resulting from a SN explosion.   The primary sources of confusion are generally H~II regions, because they are of comparable size to SNRs and can have \ha\ luminosities that easily exceed that of a typical SNR.  Planetary nebulae and nebulae around symbiotic stars are generally excluded because they are unresolved in nearby galaxies, although one LL14 candidate turned out to be a PN superposed on an H~II region.  A number of criteria can be used to distinguish SNRs from other types of nebulae (see Table \ref{snrclass}); none of the criteria used to identify nebulae as SNRs is completely definitive because unless one is observing the actual ejecta from a SN explosion, the emission that is observed is dominated by the environment into which a SN shock is expanding.  None of the methods enables one to define a complete sample of SNRs arising from SNe that exploded for a specified interval of time in the past, but some can provide high confidence of an SNR identification, especially if 2 or more criteria are met.

%


\begin{deluxetable}{llll}
\label{snrclass}
\tablecolumns{4}
\tablewidth{0pc}
\tablecaption{Proposed Criteria for SNR classification }
\tablehead{\colhead{Method} &\colhead{Criterion} &\colhead{Basis} &\colhead{Limitations}  \\
}
\startdata
[S II]/H$\alpha$  &  Greater than 0.4  & Shocks produce low ionization  & No clean separation \\
BPT Diagrams  &  Strong [O I], [O III], [N II] & Shocks produce low ionization  &  No clean separation \\
Line width   & Greater than 50 km/s  & Shock speed needed to      & Velocities reduced \\
            &                        &    produce forbidden lines &  by projection at limb \\
            Morphology    &  Shell morphology    &  Shocks at SN surface are & Confusion, complex\\
            &                        & limb-brightened  & morphology \\
X-ray emission  &  Detection  & Shocks faster than 400 km/s  & Limited sensitivity \\
            &                 & produce MK gas & \\
Radio Emission & Non-thermal emission & Shocks accelerate electrons & Limited sensitivity \\
UV emission &   Detection     & High shock temperature & Limited sensitivity, dust \\
No central star  & No hot star  & No photoionizing source & Chance superposition \\
\enddata
\end{deluxetable}

\vskip 4ex

\subsection{SNRs from the traditional [S~II]/H$\alpha$ criterion }

\begin{figure}[ht]
\includegraphics[width=4.25in]{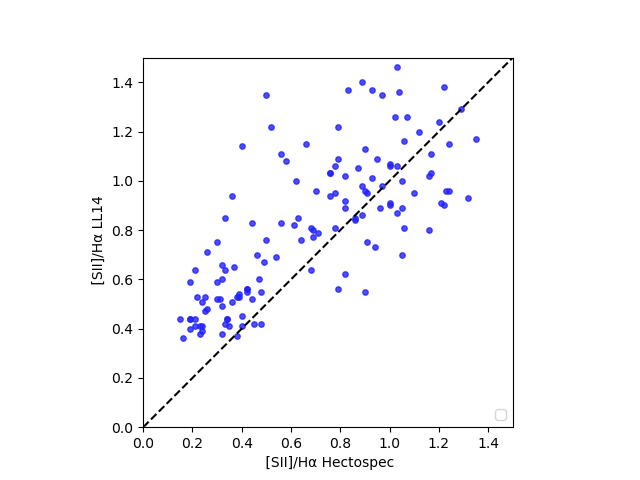} 
\caption{A comparison between the [SII]/\ha\  ratios obtained from low resolution Hectospec spectra to those obtained by \cite{lee14} from images. The dashed line shows a one-to-one correspondence.\label{fig_s2ratio_compare}}
\end{figure}

Historically, the primary method for identifying SNRs in external galaxies has been to look for identifiable nebulae with the [SII]/\ha\ ratio exceeding 0.4 ($-0.4$ in the log, which we use in  some of our plots), since H II regions, particularly higher surface brightness H II regions have [SII]/\ha\ ratios of order 0.1. While this was justified in part theoretically, the main justification was primarily observational. Indeed, this was precisely the argument made by \cite{blair81} in the first spectroscopic confirmation of SNRs in M31.   By this standard, 95 of the 135 LL14 SNR candidates we have observed spectroscopically in low resolution are SNRs.  Figure~\ref{fig_s2ratio_compare} compares the [S~II]/H$\alpha$\/ ratios listed by \citet{lee14} with those obtained from Hectospec.  Some of the scatter can be attributed to the fact that the fiber spectra pertain to a section of the nebula that might not be typical.  However, the scatter is larger than might be expected, and the [S~II]/H$\alpha$\/ ratios derived by LL14 based on images are systematically larger. As a test, we used the same images from \cite{massey} to re-derive those ratios for a small number of SNR candidates, and found that our [S~II]/H$\alpha$\/ ratios  are lower by about 0.2, and thus closer to the spectroscopic values reported here. Determination of the reason for the offset between our imaging values and those of LL14 is beyond the scope of this paper.

That a high fraction of the LL14 SNR candidates turned out to be genuine SNRs on the basis of this criterion should not be surprising.  \cite{lee14} selected nebulae on the basis of this criterion.  However, as shown in Fig.\  \ref{fig_s2ratio_compare}, the spectroscopically measured ratios tend to be lower than the optically determined values.  This is especially true of the larger diameter objects as measured by LL14.  While 77\% of the objects less than the median diameter of 45.6 pc satisfy the [SII]/\ha\ criterion, only 52\% of the larger diameter objects do. 
This is probably because shocks slower than about 100 \kmss lack the extended photoionization/recombination zone where [S~II] emission originates (see Table 1 in 
\citet{hartigan87}.

We will use the [S~II]/\ha \ ratio below (section \ref{compare})in our classification, but by necessity will need to use the line widths as well, particularly in borderline cases, or where the line ratio is absent in our data set.

\subsection{Verifying SNRs in M31 kinematically}

As noted earlier, an alternative method for identifying SNRs that is largely independent of the detailed physics of shocks is use the fact that material that radiates optically in SNRs is gas behind a shock moving at nearly the shock speed.  Shock speeds above about 40 \kmss are required to produce strong H$\alpha$, [S~II] and [N II] emission.  Even higher speeds, above about 100 \kms , are needed to produce strong [O III] emission.  Because we measure only the line of sight component of the velocity, the observed line widths can be smaller.  H II regions do have bulk motion contributions to their velocity profiles, but these are generally smaller than 30 \kms.

This method of identifying SNRs has become feasible with the availability of IFU spectrographs such as MUSE and Sitelle.   For example, \citet{li2024} have recently used MUSE to identify a large sample of SNRs and SNR candidates having  line profiles broader (after correcting for instrumental profile)  than 118 \kms\ (FWHM). For a spherical SNR, the projection onto the line of sight would mean a shock speed above 200 \kms, which would exclude many older SNRs.  On the other hand, \citet{li2024} captured the entire SNR, including the central regions where the projected speed is largest, while we may have selected emission regions near the rim of the SNR, where the projected speed is a smaller fraction of the shock speed.




\begin{figure}[ht]
\includegraphics[width=3.25in]{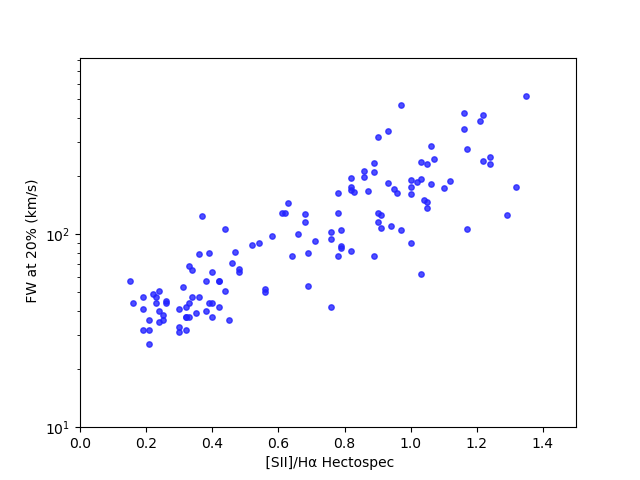} 
\caption{A comparison between the [SII]/\ha\  ratios obtained from low resolution Hectospec spectra and \ha\ line widths measured from the high resolution Hectochelle spectra.}
\label{fig:s2ratio_fwhm}
\end{figure}

\begin{figure}[ht]
\includegraphics[width=6.5in]{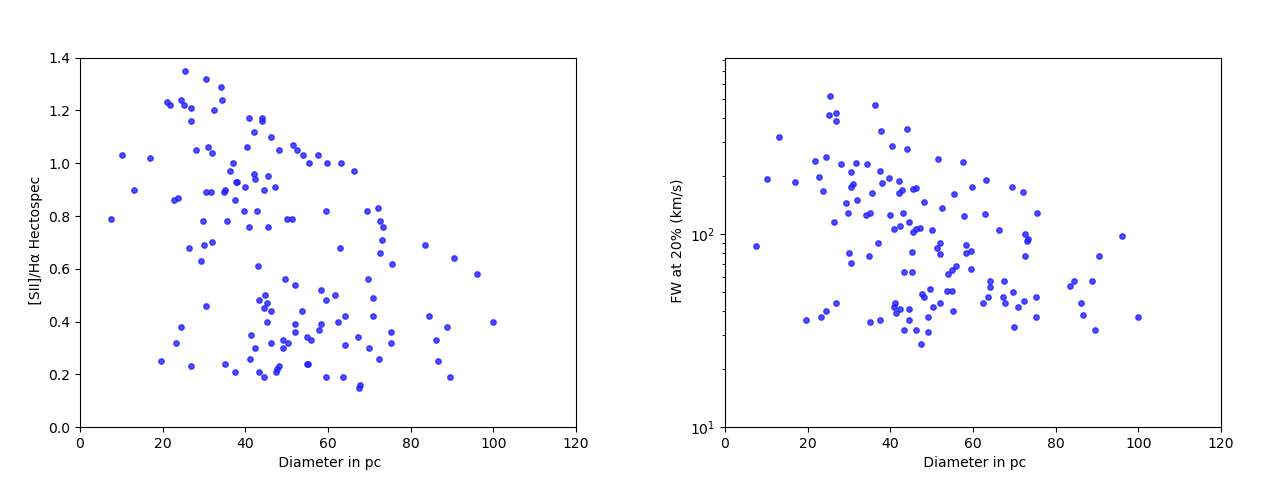} 
\caption{[SII]/\ha\ ratios (left) and velocity full width at 20\% of maximum line intensity (right) as a function of object diameter,  the latter taken from table 3 of LL14}.  In both cases, there is more dispersion in the values at small diameters than at large diameters.
\label{fig:dia}
\end{figure}

As shown in Fig.\  \ref{fig:s2ratio_fwhm}, there is a very strong correlation between [SII]/\ha\/ line ratios and velocity broadening in \ha. The models of \citet{hartigan87} indicate that while high [SII]/\ha\ ratios can occur in extremely slow shocks, the ratio is less than 0.4 in shocks below 100 \kms, and it increases to values larger than 1.0 at 250 \kms.  Diameters measured in  the [SII] images were presented by LL14, which we present in Figure\  \ref{fig:dia} showing that candidates with high [SII]/\ha\ ratios  and large velocity FWHM tend to be smaller diameter objects.  This is what one might expect if shocks speeds slow as a SNR ages.  The large spread of ratios and FWHM even at small diameters can be interpreted as reflecting the fact that SNRs in high density environment evolve more rapidly.

 We therefore take line widths above 50 \kmss (FW$_{20}$) to be a strong indication that the nebula is an SNR.  Note that many of the line profiles are not Gaussian.   Aside from various skewed profiles, there are cases where a very narrow line core sits on top of a broader component.  This is most easily interpreted as the superposition of SNR emission on emission from an H II region that may or may not be physically close by.  In specific, a nebula with a width greater than 65 \kms \ will be considered a certain SNR, while those with widths between 50 and 65 \kms \ will be considered possible SNRs, unless the [SII]/\ha\ ratio is also larger than 0.6, providing more certitude.
 
\begin{figure}[ht]
\includegraphics[width=3.5in]{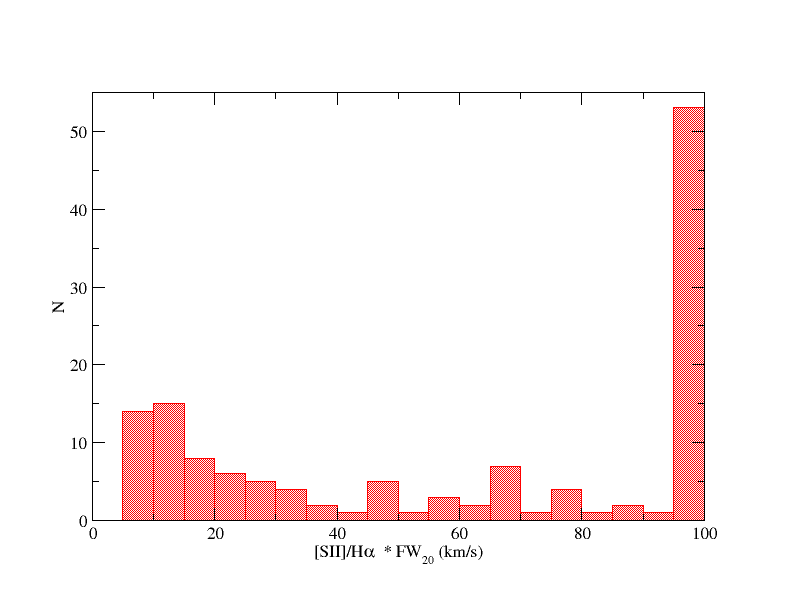} 
\caption{ A histogram of the product of [SII]/\ha\ ratio and the velocity width (full-width at 20\% of the maximum height in this case). We lumped also values above 100 in the group at 100. HII regions clearly have the minimum values, but there is no clear gap in values between those and the obvious SNRs at the highest values> However, we set the classification of objects in Table \ref{table3} with values below 15 to by HII regions, and those above 20 to be SNR. In between those values we set as SNR* or Uncertain.}
\label{fig:s2_fwhm}
\end{figure}

\cite{duarte-puertas24} advocated for using the product of the [SII]/\ha\ ratio and the line width to distinguish HII regions and SNRs in their study of emission nebulae in M33. We present Fig \ref{fig:s2_fwhm} which shows the M31 sample (115 in this plot), where we have used the \ha \   (FW$_{20}$), and set the maximum value to be 100 for visual purposes. Clearly, HII regions have the lowest values, and the most obvious SNRs have values larger than 100, but without a clear demarcation. For Table \ref{table3}, where needed we set objects with values less than 15 to be HII regions, and those greater than 20 to be SNR.

\subsection{Verifying SNRs through their X-ray, Radio and UV Emission}


X-ray emission, non-thermal radio emission, and strong UV emission can all be indications of shock heating.  In Table \ref{table3}, we list the 50 objects in our sample that have been detected at these wavelengths, along with the morphological classification of LL14.  The following sections discuss the details of emission at other wavelengths.

\subsubsection{X-rays}

Supernova remnants typically produce both X-ray emission from shocks faster than about 400 \kms\/ and optical emission from slower shocks.  The slower shocks occur in denser material, and the ram pressures of the fast and slow shocks are related by roughly equal ram pressures, $\rho V^2$.  The X-ray emission is generally dominated by optically thin thermal emission in the 0.2 $<~kT~< 3 ~\rm keV$ range, and detection of such emission in the absence of a central point source is a very strong criterion for the identification of an SNR.  

One limitation is that existing X-ray surveys have limited sensitivity and spatial resolution, though eRosita should help to remedy this.  Also, the X-ray emission is faint in SNRs in low density ambient gas (galactic halos or superbubbles) and in old SNRs at the low end of the temperature range, especially if there is significant absorption by intervening gas.  The other problem is that many of the observed X-ray sources are X-ray binaries or AGN.  Most of these other sources have harder spectra than SNRs, but Supersoft Sources could potentially be confused with SNRs

Some of the objects identified as SNR from the [S~II]/H$\alpha$ ratio are within 10\arcsec \ of an X-ray source from \cite{stiele11}.  Several recent papers on X-ray sources in M31 provide additional detections of the \citet{lee14} SNR candidates \citep{sasaki, stiele11, vulic, williams18, huang23},  and we rely on those identifications.  Of the 152 candidates from \citet{lee14} for which we have optical data, 30 are detected in X-rays, and all except one of those 30 show either high [S~II]/H$\alpha$ ratios or wide line profiles or both. The one exception has no current [S~II]/H$\alpha$ measure, and a  line-width below our SNR cutoff (LL14-131).  

\subsubsection{Radio}

Non-thermal radio emission is a clear signature of a shock.  It can be produced by energetic electrons accelerated by the shock or by ambient energetic electrons
reaccelerated in the compressed ambient magnetic field \citep{raymond2020, tutone2021}.  \citet{galvin2014} measured fluxes at two wavelengths for 98 targets in M31, though they were only able to measure spectral indices for 9.  Of those, they found that 73\% were steeper than $\alpha$ = -0.6 and therefore nonthermal.  These could be SNRs or background sources.  They compared their full detection list, including those without spectral indices, with SNR candidates and found that only 14 of the SNR candidates showed radio emission.  All the sources they reported that we observed satisfied the [S~II] and line width criteria.  Eight of their sources were detected in X-rays and six were not.  The two main drawbacks of using radio emission as a criterion for SNRs are the limited sensitivity of existing radio studies and contamination by bright thermal emission from H II regions in nearby star-forming regions.  Unfortunately, only two of the \citet{galvin2014} sources that we observed (LL14-20 and LL14-70) had measured spectral indices that proved a non-thermal spectrum and confirmed that those objects are SNRs

\subsubsection{UV}

H II regions are generally cooler than $10^4$ K (kT $<$ 1 eV), while SNR shocks produce temperatures of order $10^5$ K or more (kT $>$ 10 eV).  Since optical lines have excitation energies around 2 eV, while UV lines such as C IV $\lambda$1550 have excitation energies around 8 eV, the Boltzmann factor in the excitation rates implies much higher UV/optical line ratios in shocks than in photoionized gas.  Over the lifetime of an SNR, a significant fraction of SN luminosity emerges in the UV \citep{makarenko23}.  The main limitations of this criterion are the general lack of high-sensitivity UV surveys and the severe reduction of UV fluxes by dust. The UV continuum is also a way to determine whether or not there is a hot star that can photoionize the nebula. 

Recently, \cite{leahy2023} searched the sites of the candidate SNRs in images of M31 obtained with UV Imaging Telescope (UVIT) on ASTROSAT. Of approximately 40 SNR candidates selected from optical, radio and X-ray lists, they report that 20 candidates were detected.  The UV luminosities reported are comparable to those of a sample of Galactic and LMC SNRs.  \citet{leahy2023} suggest that the reason they do not detect more SNRs is that some SNRs lie behind substantial interstellar absorption within M31 itself.

 All of the 19 objects from the \citet{lee14} catalog reported as detected with UVIT are in our sample, and of those, 18 were observed with Hectochelle.  Among those 18, 15 of the objects showed FW$_{20} >$ 50 \kms, confirming that they are SNRs. 
Two objects (LL14-130 and -147) have line widths well below 50 \kms, indicating they are HII regions (the [S~II]/\ha\ ratio confirms this for LL14-147). LL-135 is an ambiguous case, with a width of 50 \kms \ but a high [S~II]/\ha \  ratio. However, there is little correlation between strength of [S~II]/\ha\  or line width and UVIT detection. Overall, this suggests that UV emission, like the other criteria, is a useful but not foolproof discriminant.

\subsection{Comparing Criteria from Different Wavelengths \label{compare}}

\begin{figure}[ht]
\begin{center}
\includegraphics[width=3.25in]{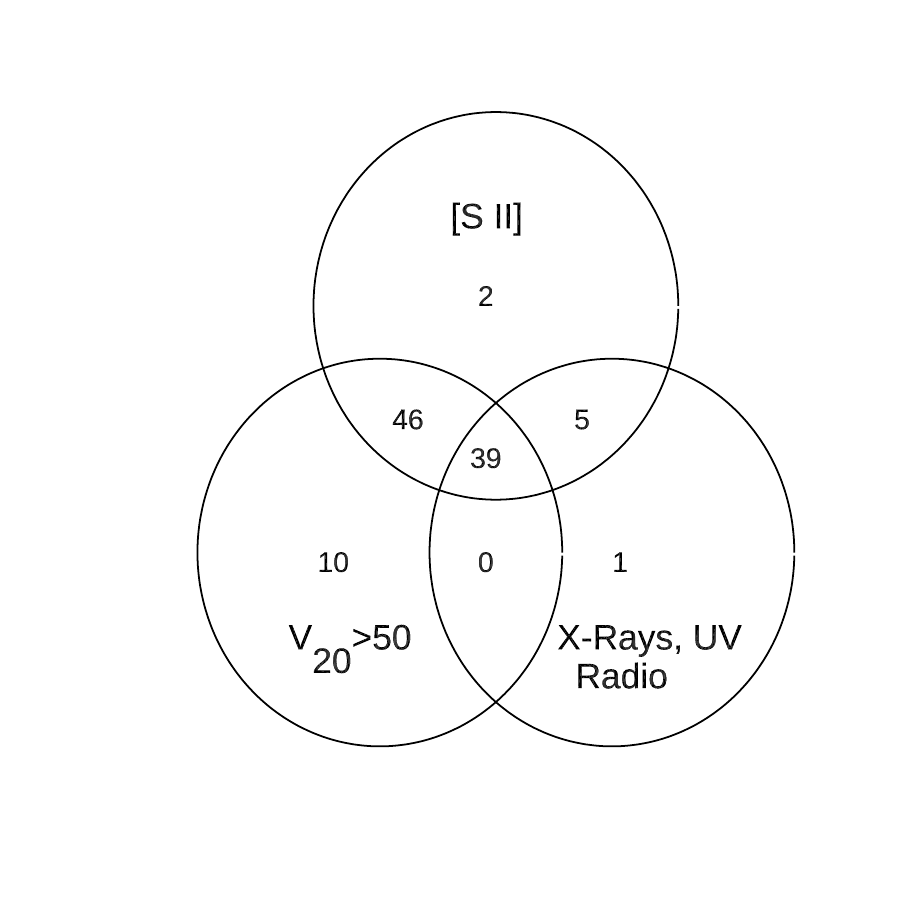} 
\caption{ Venn diagram of the [S~II], line width criteria and detection at X-ray,
radio or UV wave lengths for the 99 objects with full data. Of the 152 LL14 objects in our sample, 22 had only one optical criteria, and 33 more failed to meet either the [S~II] or FW$_{20}$ criteria, and they are not included here.}
\label{fig:venn}
\end{center}
\end{figure}

To demonstrate capabilities and limitations of the different methods of identifying SNRs, Figure~\ref{fig:venn} shows a Venn diagram for objects from the \citet{lee14} candidates that satisfy the criteria [S~II]/H$\alpha~>~0.4$, FW$_{20}$ $>$ 50 \kms, or detection at X-ray, radio and/or UV wavelengths.  The X-ray catalog of \citet{huang23} includes 27 objects in our sample.  The radio catalog of \citet{galvin2014} contains 13 sources within 5\arcsec\ of an object in our sample, and \citet{leahy2023} lists 14 of the objects we observed.  We have low resolution data for 135 sources and high resolution line profiles for 147, giving us a total of 152 for which we have at least one kind of spectrum and 130 which have both.  Note that all of the \citet{lee14} candidates were chosen on the basis of the [S~II]/H$\alpha$\ ratio measured using imaging data, and there are 21 for which we do not have low resolution data. Of the 135 candidates for which we have low resolution spectra, 88 satisfy the [S~II] criterion, meaning that 47 objects which according to \citet{lee14} satisfy the criterion based on the imaging data, do not satisfy the criterion according to the spectra.  Possible reasons for the disagreement are inaccuracy of the images derived from the images, contributions from the Diffuse Interstellar Gas, or variations within the nebulae such that Hectospec fiber happened to be placed on a low [S~II]/H$\alpha$\ region.  Some regions are probably SNRs based on the line width or X-ray criteria, but there are 32 candidates for which we have both optical criteria, and neither is satisfied, so they are probably H II regions. There are another 4 for which we have just one optical measurement, and that one measurement alone clearly rules out the object being an SNR .

 In summary, of the 152 objects in our sample of \citet{lee14} supernova remnant candidates, 39 satisfy both optical criteria and are also detected in X-rays, radio and/or UV. For 22 objects we had high- or low-resolution spectra, but not both.  Of those 22, 12 had [S~II]/H$\alpha$ larger than 0.6 or FW$_{20}$ larger than 65 \kms, and we consider those values to be above the thresholds by more than the uncertainties, and therefore to be SNRs, for a total of 104 confirmed SNRs out of our sample of 152 LL14 SNR candidates, with 5 more that don't have complete optical data.  

Table \ref{table3} lists the X-ray, UV and radio detections along with the classification for each LL14 source.  The [S~II]/\ha \ ratios and the FW$_{20}$ line widths are taken from Table \ref{ll14_spectra}. The X-ray fluxes from \citet{huang23} are in units of $10^{-14}~\rm erg~cm^{-2}~s^{-1}$, the UV luminosities from \citet{leahy2023} in the F148W band are in units of $10^{35}~\rm erg~s^{-1}$, and the radio fluxes at 1.4 GHz from \citet{galvin2014} are in mJy. The classification as supernova remnant, H II region or Uncertain is based on the two optical criteria in the previous paragraphs, summarized in Table \ref{snrclass_final}. Complications are due to incomplete data for some objects. There are 10 objects where one optical criterion indicates an SNR, but the other either is missing or at the margin of that limit. For those cases we use the product of the ratio and widths as above to help provide a distinct dividing line, albeit artificial.  We place an asterisk on the objects where one criterion indicates an SNR, but the other is absent, "SNR*."  Other intermediate objects we denote as "Uncertain." 
\begin{deluxetable}{lllr}
\tablecolumns{4}
\tablewidth{0pc}
\tablecaption{Detailed Optical Criteria for SNR classification\tablenotemark{a}}
\tablehead{\colhead{Level}  &\colhead{Criteria} &\colhead{Type} &\colhead{Number}\\
}
\startdata
(a) & [S~II]/H$\alpha > 0.6\ $ {\it or} FW$_{20}>=65$ & SNR &97\\\
(b) &no [S~II]/H$\alpha $ {\it and} $50 < $ FW$_{20}<=65$  & SNR* & 1\\\
(c) &$0.4 <=$ [S~II]/H$\alpha $$< 0.6\ $ {\it and}  no FW$_{20}$  & SNR* & 2\\\
(d) & [S~II]/H$\alpha * $ FW$_{20} >= 20 $ {\it and} not (a) & SNR* & 9\\\ 
(e) &[S~II]/H$\alpha <0.3$ {\it or}  FW$_{20} <30$ & HII &20\\\
(f) & [S~II]/H$\alpha * $ FW$_{20} > 15$ {\it and} not also (e)   & HII & 9\\\ 
(g) &no [S~II]/H$\alpha $ \ {\it and} FW$_{20}< 50$  & Uncertain & 5\\\
(h) & $15 <= $[S~II]/H$\alpha * $ FW$_{20} < 20 $ & Uncertain & 8\\\ 
\enddata
\tablenotetext{a}{Levels (g) and (h), "Uncertain," fill in the gaps of the other levels}
\label{snrclass_final}
\end{deluxetable}

\newpage
\startlongtable
\begin{deluxetable}{rrrrrrr|rrrrrrr}
\tabletypesize{\scriptsize}
\decimals
\tablecaption{Optical, X-ray, Radio and UV Properties of LL14 Objects}
\tablehead{
\colhead{Source} &
\colhead{S II/H$\alpha$} &
\colhead{FW$_{20}$} &
\colhead{X-ray} &
\colhead{Radio} &
\colhead{UV} &
\colhead{Type} &
\colhead{Source} &
\colhead{S II/H$\alpha$} &
\colhead{FW$_{20}$} &
\colhead{X-ray} &
\colhead{Radio} &
\colhead{UV} &
\colhead{Type}  \\
\colhead{}   &\colhead{}  &\colhead{km/s} &\colhead{F$_{-14}$}  &\colhead{mJy} &{L$_{35}$} &\colhead{} &\colhead{} &\colhead{}  &\colhead{km/s} &\colhead{F$_{-14}$}  &\colhead{mJy} &{L$_{35}$} &\colhead{}\\
}
\startdata
LL14-001	&	--	&	205	&	--	&	--	&	--	&	SNR & LL14-077	&	0.3342	&	37	&	--	&	--	&	--	&	HII \\  
LL14-002	&	1.0046	&	162	&	3.8	&	--	&	--	&	SNR & LL14-078	&	0.9268	&	341	&	--	&	--	&	17.0	&	SNR \\  
LL14-003	&	0.6966	&	83	&	--	&	--	&	--	&	SNR & LL14-079	&	0.6353	&	77	&	--	&	--	&	40.0	&	SNR \\  
LL14-004	&	0.9484	&	171	&	0.3	&	--	&	--	&	SNR & LL14-080	&	1.1615	&	351	&	1.3	&	1.1	&	--	&	SNR \\  
LL14-005	&	--	&	46	&	--	&	--	&	--	&	Unc & LL14-081	&	0.3119	&	53	&	--	&	--	&	--	&	Unc \\  
LL14-006	&	0.5420	&	54	&	--	&	0.7	&	50.0	&	SNR & LL14-082	&	0.8974	&	318	&	--	&	--	&	6.2	&	SNR \\  
LL14-007	&	0.8260	&	165	&	--	&	--	&	--	&	SNR & LL14-083	&	0.4808	&	66	&	3.7	&	--	&	--	&	SNR \\  
LL14-008	&	1.0691	&	245	&	--	&	--	&	--	&	SNR & LL14-084	&	0.3573	&	79	&	--	&	--	&	--	&	SNR \\  
LL14-009	&	0.7586	&	86	&	--	&	--	&	--	&	SNR & LL14-085	&	--	&	102	&	--	&	--	&	9.9	&	SNR \\  
LL14-010	&	0.7907	&	87	&	--	&	--	&	--	&	SNR & LL14-086	&	0.2208	&	41	&	--	&	--	&	--	&	HII \\  
LL14-011	&	0.3767	&	40	&	--	&	--	&	--	&	Unc & LL14-087	&	1.2274	&	--	&	2.1	&	1.9	&	--	&	SNR \\  
LL14-012	&	0.7889	&	85	&	--	&	--	&	--	&	SNR & LL14-088	&	1.1194	&	189	&	--	&	--	&	--	&	SNR \\  
LL14-013	&	0.2388	&	40	&	--	&	--	&	--	&	HII & LL14-089	&	0.3304	&	68	&	--	&	--	&	--	&	SNR \\  
LL14-014	&	0.1866	&	--	&	--	&	--	&	--	&	pn &  LL14-090	&	0.3243	&	37	&	--	&	--	&	--	&	HII \\  
LL14-015	&	0.1932	&	47	&	--	&	--	&	--	&	HII & LL14-091	&	--	&	57	&	--	&	--	&	--	&	SNR* \\ 
LL14-016	&	0.3990	&	44	&	--	&	--	&	--	&	Unc & LL14-092	&	0.3606	&	47	&	--	&	--	&	--	&	Unc \\  
LL14-017	&	0.2427	&	35	&	--	&	--	&	--	&	HII & LL14-093	&	0.4989	&	--	&	--	&	--	&	14.0	&	SNR* \\ 
LL14-018	&	1.3521	&	518	&	2.1	&	2.6	&	--	&	SNR & LL14-094	&	0.8974	&	128	&	--	&	--	&	--	&	SNR \\  
LL14-019	&	0.2312	&	44	&	--	&	--	&	--	&	HII & LL14-095	&	0.2449	&	51	&	--	&	--	&	--	&	HII \\  
LL14-020	&	1.2246	&	240	&	--	&	2.5	&	--	&	SNR & LL14-096	&	0.3436	&	65	&	--	&	--	&	--	&	SNR \\  
LL14-021	&	1.0520	&	231	&	--	&	--	&	--	&	SNR & LL14-097	&	0.9141	&	107	&	--	&	--	&	--	&	SNR \\  
LL14-022	&	0.2089	&	36	&	--	&	--	&	--	&	HII & LL14-098	&	0.8222	&	175	&	--	&	--	&	86.0	&	SNR \\  
LL14-023	&	0.4246	&	42	&	--	&	--	&	--	&	Unc & LL14-100	&	--	&	123	&	4.4	&	--	&	--	&	SNR \\  
LL14-024	&	1.0471	&	146	&	--	&	--	&	--	&	SNR & LL14-101	&	0.7586	&	102	&	--	&	--	&	--	&	SNR \\  
LL14-025	&	0.2979	&	31	&	--	&	--	&	--	&	HII & LL14-102	&	0.5383	&	90	&	--	&	--	&	--	&	SNR \\  
LL14-026	&	--	&	65	&	--	&	--	&	--	&	SNR & LL14-103	&	0.6792	&	127	&	--	&	--	&	--	&	SNR \\  
LL14-027	&	--	&	93	&	--	&	--	&	--	&	SNR & LL14-104	&	0.7907	&	105	&	--	&	--	&	--	&	SNR \\  
LL14-028	&	--	&	38	&	--	&	--	&	--	&	Unc & LL14-105	&	1.0328	&	192	&	0.8	&	--	&	--	&	SNR \\  
LL14-029	&	0.7834	&	77	&	--	&	--	&	--	&	SNR & LL14-106	&	0.9638	&	164	&	1.0	&	--	&	38.0	&	SNR \\  
LL14-030	&	0.6166	&	129	&	--	&	--	&	--	&	SNR & LL14-107	&	--	&	116	&	0.2	&	--	&	--	&	SNR \\  
LL14-031	&	1.1722	&	106	&	--	&	--	&	--	&	SNR & LL14-108	&	0.8913	&	209	&	1.3	&	--	&	--	&	SNR \\  
LL14-032	&	1.0280	&	237	&	--	&	2.2	&	--	&	SNR & LL14-109	&	0.8166	&	169	&	0.6	&	--	&	--	&	SNR \\  
LL14-033	&	0.4457	&	36	&	--	&	--	&	--	&	Unc & LL14-110	&	0.1560	&	44	&	--	&	--	&	--	&	HII \\  
LL14-034	&	1.1615	&	424	&	5.4	&	0.7	&	--	&	SNR & LL14-111	&	0.8851	&	232	&	1.9	&	--	&	--	&	SNR \\  
LL14-035	&	0.3516	&	39	&	--	&	--	&	--	&	HII & LL14-112	&	1.2388	&	252	&	1.1	&	1.5	&	--	&	SNR \\  
LL14-036	&	0.8204	&	82	&	--	&	--	&	--	&	SNR & LL14-113	&	0.6123	&	129	&	--	&	--	&	76.0	&	SNR \\  
LL14-037	&	1.1749	&	276	&	1.2	&	--	&	--	&	SNR & LL14-114	&	0.7816	&	129	&	0.7	&	--	&	--	&	SNR \\  
LL14-038	&	1.0233	&	187	&	--	&	--	&	--	&	SNR & LL14-115	&	0.6823	&	116	&	--	&	--	&	--	&	SNR \\  
LL14-039	&	1.2023	&	--	&	0.3	&	--	&	--	&	SNR & LL14-116	&	0.2123	&	32	&	--	&	--	&	--	&	HII \\  
LL14-040	&	0.4656	&	81	&	--	&	--	&	49.0	&	SNR & LL14-117	&	0.3006	&	33	&	--	&	--	&	--	&	HII \\  
LL14-041	&	1.2388	&	230	&	1.7	&	--	&	--	&	SNR & LL14-118	&	0.8913	&	77	&	--	&	--	&	--	&	SNR \\  
LL14-042	&	0.9311	&	184	&	--	&	--	&	--	&	SNR & LL14-119	&	0.2323	&	47	&	--	&	--	&	--	&	HII \\  
LL14-043	&	--	&	186	&	--	&	--	&	--	&	SNR & LL14-120	&	0.4395	&	106	&	--	&	--	&	--	&	SNR \\  
LL14-044	&	0.4560	&	71	&	--	&	--	&	--	&	SNR & LL14-121	&	--	&	40	&	--	&	--	&	--	&	Unc \\  
LL14-045	&	0.8630	&	211	&	--	&	--	&	--	&	SNR & LL14-122	&	0.2630	&	45	&	--	&	--	&	--	&	HII \\  
LL14-046	&	0.6281	&	145	&	--	&	--	&	--	&	SNR & LL14-123	&	1.0351	&	150	&	--	&	--	&	--	&	SNR \\  
LL14-047	&	1.0520	&	137	&	1.5	&	--	&	--	&	SNR & LL14-124	&	0.9727	&	467	&	2.4	&	2.6	&	--	&	SNR \\  
LL14-048	&	1.3183	&	176	&	2.8	&	1.5	&	--	&	SNR & LL14-125	&	0.5649	&	50	&	--	&	--	&	14.0	&	SNR \\  
LL14-049	&	0.3715	&	124	&	--	&	--	&	--	&	SNR & LL14-126	&	0.3155	&	37	&	--	&	--	&	--	&	HII \\  
LL14-050	&	1.2078	&	384	&	4.1	&	--	&	47.0	&	SNR & LL14-127	&	0.3350	&	44	&	--	&	--	&	--	&	HII \\  
LL14-051	&	0.9376	&	110	&	1.0	&	--	&	--	&	SNR & LL14-128	&	1.0280	&	62	&	--	&	--	&	--	&	SNR \\  
LL14-052	&	1.0023	&	175	&	--	&	--	&	8.1	&	SNR & LL14-129	&	0.8610	&	198	&	--	&	--	&	--	&	SNR \\  
LL14-053	&	0.6561	&	100	&	--	&	--	&	--	&	SNR & LL14-130	&	--	&	34	&	--	&	--	&	22.0	&	Unc \\  
LL14-054	&	0.7129	&	92	&	--	&	--	&	--	&	SNR & LL14-131	&	--	&	39	&	2.1	&	--	&	--	&	Unc \\  
LL14-055	&	0.3908	&	80	&	--	&	--	&	--	&	SNR & LL14-132	&	0.2535	&	36	&	--	&	--	&	--	&	HII \\  
LL14-056	&	0.4207	&	57	&	--	&	--	&	--	&	SNR & LL14-133	&	0.8710	&	167	&	--	&	2.1	&	--	&	SNR \\  
LL14-057	&	0.5176	&	88	&	--	&	--	&	--	&	SNR & LL14-134	&	0.1500	&	57	&	--	&	--	&	--	&	HII \\  
LL14-058	&	0.9954	&	90	&	--	&	--	&	--	&	SNR & LL14-135	&	0.3420	&	47	&	--	&	--	&	--	&	Unc \\  
LL14-059	&	0.9141	&	126	&	1.5	&	--	&	--	&	SNR & LL14-136	&	0.4753	&	64	&	--	&	--	&	--	&	SNR \\  
LL14-060	&	0.3954	&	64	&	--	&	--	&	--	&	SNR* &LL14-137	&	0.9727	&	105	&	--	&	--	&	20	&	SNR \\  
LL14-061	&	1.1015	&	174	&	1.0	&	--	&	--	&	SNR & LL14-138	&	0.3908	&	44	&	--	&	--	&	--	&	Unc \\  
LL14-062	&	0.2535	&	38	&	--	&	--	&	--	&	HII & LL14-139	&	0.7798	&	163	&	--	&	--	&	--	&	SNR \\  
LL14-063	&	0.2178	&	49	&	--	&	--	&	--	&	HII & LL14-140	&	0.5754	&	98	&	--	&	--	&	--	&	SNR \\  
LL14-064	&	0.9954	&	191	&	--	&	--	&	--	&	SNR & LL14-141	&	0.3184	&	42	&	--	&	--	&	--	&	HII \\  
LL14-065	&	--	&	75	&	--	&	--	&	--	&	SNR & LL14-142	&	0.4385	&	51	&	--	&	--	&	--	&	SNR \\  
LL14-066	&	1.0568	&	285	&	9.1	&	--	&	--	&	SNR & LL14-143	&	0.8954	&	116	&	--	&	1.2	&	--	&	SNR \\  
LL14-067	&	--	&	76	&	--	&	--	&	11	&	SNR & LL14-144	&	0.6887	&	54	&	--	&	--	&	--	&	SNR \\  
LL14-068	&	--	&	80	&	--	&	--	&	--	&	SNR & LL14-145	&	0.3846	&	57	&	--	&	--	&	--	&	SNR* \\ 
LL14-069	&	0.3034	&	41	&	--	&	--	&	--	&	HII & LL14-146	&	0.3999	&	37	&	--	&	--	&	--	&	HII \\  
LL14-070	&	1.2218	&	412	&	3.6	&	1.0	&	44.0	&	SNR & LL14-147	&	0.1950	&	32	&	--	&	--	&	28.0	&	HII \\  
LL14-071	&	1.0641	&	181	&	--	&	--	&	--	&	SNR & LL14-148	&	0.5623	&	52	&	--	&	--	&	28.0	&	SNR \\  
LL14-072	&	0.2148	&	27	&	--	&	--	&	--	&	HII & LL14-149	&	0.3162	&	32	&	--	&	--	&	--	&	HII \\  
LL14-073	&	1.2942	&	126	&	--	&	--	&	--	&	SNR & LL14-150	&	0.4246	&	57	&	--	&	--	&	--	&	SNR \\  
LL14-074	&	0.2576	&	44	&	--	&	--	&	--	&	HII & LL14-152	&	0.5035	&	--	&	--	&	--	&	--	&	SNR* \\ 
LL14-075	&	0.6855	&	80	&	--	&	--	&	--	&	SNR & LL14-154	&	--	&	232	&	--	&	--	&	--	&	SNR \\  
LL14-076	&	0.8166	&	194	&	1.7	&	--	&	--	&	SNR & LL14-155	&	0.7586	&	94	&	--	&	--	&	--	&	SNR \\   
\enddata
\label{table3}
\end{deluxetable}

\subsection{Other Criteria}

A number of other criteria have been used to discriminate among SNRs, H II regions and planetary nebulae.  We briefly discuss the rationales for and the limitations of morphology and the use of other emission lines. 



\subsubsection{Other emission lines}

\cite{kopsacheili20} among others have argued that a more robust way to identify SNRs is through a combination of line fluxes, with increased emphasis on [O~I], which is expected to be strong in complete shocks, but weak in H~II regions where most of O is in higher ionization states.  The arguments they make are largely based on model calculations made using MAPPINGS. However, to our knowledge, it has not been verified that a Galactic or Magellanic Cloud SNR exists (where the SNR identity is fully established), that does not also satisfy the [S~II]/\ha\ criterion.

Optical identification of SNRs from spectra or from narrow-band images is often based on the criterion [S~II]/H$\alpha > 0.4$.  The basis for this criterion is that elements such as O and S are generally doubly ionized in photoionized nebula.  On the other hand, the cooling region behind a shock usually has a recombination/photoionization zone at a temperature below $10^4$ K, where singly ionized species such as S~II, N II and O II dominate and O I is also abundant.  

\citet{li2024} used the [S~II]/H$\alpha$ and [O I]/H$\alpha$ ratios, along with BPT diagrams and [S~II] line widths, to identify SNRs in MUSE IFU observations of 19 galaxies.  They found that [O I]/H$\alpha$ $>$ 0.1 is the most successful selection criterion in the sense that all targets above that ratio can be considered to be SNRs.  However, many SNR as identified via other criteria can be missed by using that ratio exclusively.  \citet{li2024} also used residual [S~II]/H$\alpha$ ratio, subtracting off the expected contribution from diffuse emission and H II regions based on fits to the relation between [S~II]/H$\alpha$ vs H$\alpha$ flux for their galaxies.  We expect these contributions to be relatively smaller for our M31 sample because our 1.5\arcsec\/ fibers correspond to about 5.5 pc in M31.  On the other hand, the \citet{li2024} resolution corresponds to 67 pc on average, which is larger than the SNRs being studied, and their spectra therefore include more foreground and background emission.

\begin{figure}[ht]
\includegraphics[width=7.25in]{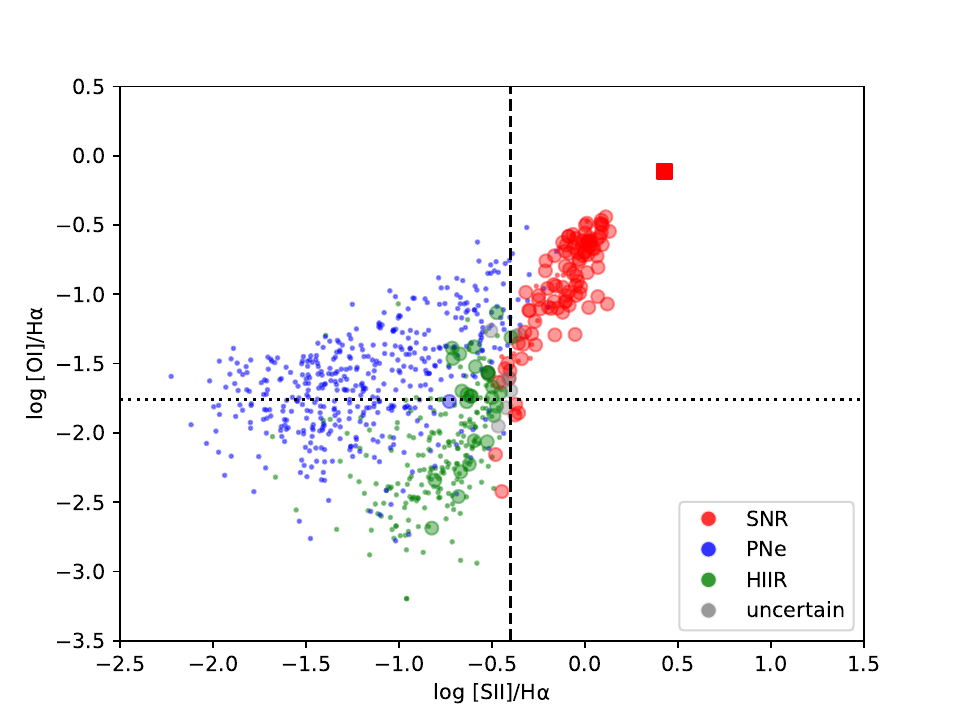} 
\caption{A [OI]/\ha \ versus [SII]/\ha \ BPT diagram of emission-line objects in M31 observed with Hectospec. Large points are the LL14 objects. Smaller points are PNe and HII regions published in earlier papers cited in the text, where their classifications come from their location in the [OIII]/\hbeta \ versus [NII]/\ha \ BPT diagram, as described in \cite{SCM} and \cite{bhatt19}.   The dashed vertical line denotes the [S~II]/H$\alpha$ ratio of 0.4 criterion separating SNR from HII regions, while the horizontal dotted line is the value of [O~I]/\ha \ used by \cite{kopsacheili20} in a similar way. The young, nitrogen-rich SNR WB92-26 is shown as a large square. Note that a number of the LL14 SNR candidates (large symbols) have  [SII]/\ha \ ratios below $0.4$  ($-0.4$ in the log), meaning those were HII regions misidentified as SNRs.
}
\label{fig:bpt_o1}
\end{figure}

The sample of nebulae selected by \cite{lee14} were all based on the [SII]/\ha\  criterion (even though it is clear from our spectroscopic data that not all of them actually are SNRs).  As a result, we cannot comment on the possibility that there are a large number of nebulae that are likely to be SNRs that would have been identified if nebula with significant [O~I]/\ha\ were selected. That said, it is worthwhile to ask, how many of the nebulae that have been suggested to be SNRs in M31  have  anomalously high [O I]/\ha\ ratios, but do not satisfy the historical [SII]/\ha\ criterion. To gain some insight into this possibility, Fig.~\ref{fig:bpt_o1} shows the line ratio of [OI]/\ha\ as a function of the [SII]/\ha\ ratio.  For comparison, the plot includes spectra from the H II region and PNe samples mentioned above, along with the full \citet{lee14} SNR sample. There is evidently a strong correlation between the [SII]/\ha\ line ratio and the ratio of [O~I]/\ha.  
The correlations that we do observe are consistent with what one would expect if ratios increase with shock velocity, as is the case observationally and as was discussed by \cite{winkler23} for the case of M83 (see Fig.\  6 of that paper).

\begin{figure}[ht]
\includegraphics[width=7.25in]{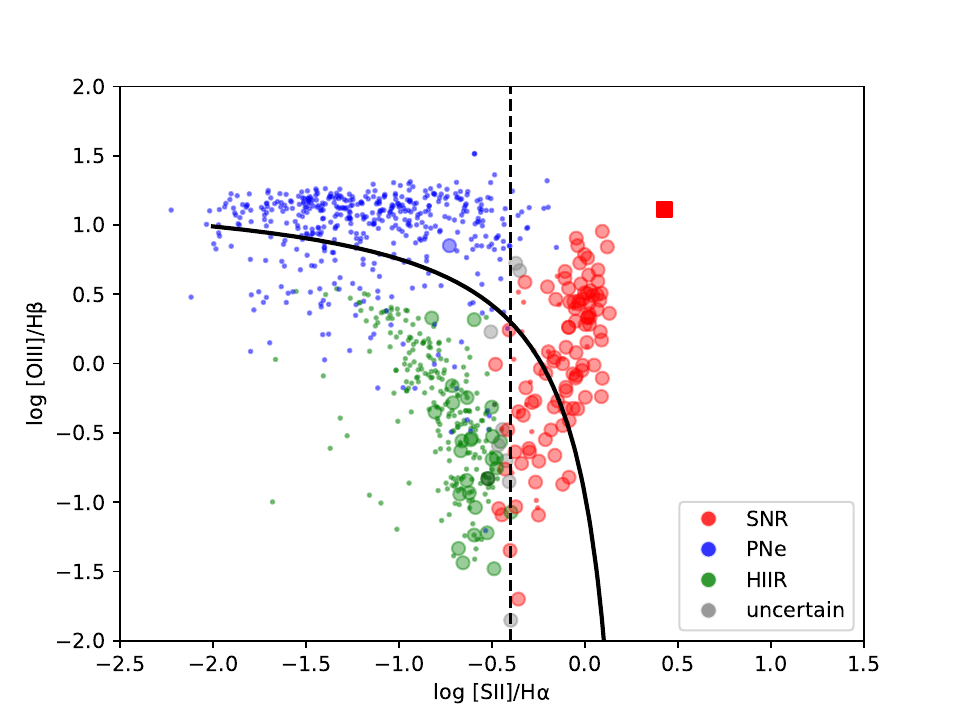} 
\caption{The [SII]/\ha \ versus [OIII]/\hbeta \ BPT diagram, for objects reported here or observed with Hectospec and published in the papers described in the text. Colors are as above and large points are the LL14 objects.  The dashed line denotes the [S~II]/H$\alpha$ ratio of 0.4 criterion. The solid line denotes the separation between shocked and photoionized gas that \citet{kewley06} presented for AGN spectra.  Note that PNe would not be confused with SNRs in M31 because they would not be resolved, though one chance superposition is evident - see the larger LL14 symbol in the PNe grouping. PNe with log [S~II]/H$\alpha > -0.4$ are those low-density PNe with very high [NII]/H$\alpha$. The young, nitrogen-rich SNR WB92-26 is again shown as a large square.}
\label{fig:bpt_s2c}
\end{figure}

Another ratio-ratio diagram that is widely used shows [O III]/H$\beta$\/ against [S~II]/H$\alpha$.  The separation between SNRs and other nebulae is often drawn according to models of shock waves and photoionized regions, for instance in \citet{li2024}.  Figure~\ref{fig:bpt_s2c} shows that there is no sharp distinction between SNRs and other H II regions, though PNe are fairly separate.  Among the objects in our sample for which we have high resolution spectra, 55 lie above both the \citet{kewley06} curve and [S~II]/H$\alpha$ = 0.4 line.  Though this curve was meant to differentiate between H II regions on the one hand and LINERS/Seyfert galaxies on the other, it is sometimes taken as an indicator of photoionized vs. shock-excited emission.  Of those, 53 have line widths FW$_{20 }>$ 50 \kms, and they are clearly SNRs, and the other two have [S~II]/H$\alpha$ ratios barely above 0.4.  In the region below the \citet{kewley06} curve but above the [S~II]/H$\alpha$ = 0.4 line, there are 27 objects, of which 23 have FW$_{20}$ larger than 50 \kms, 2 have FW$_{20}$  less than 50 \kms, and 2 are marginal at 50 \kms.  We conclude that nearly all the objects in that sector are SNRs.  There are 42 objects in the region below the \citet{kewley06} curve and below the [S~II]/H$\alpha$ = 0.4 line, of which 26 show FW20 below 50 \kms, 14 show FW$_{20}$ above 50 \kms, and 2 are marginal.  Thus about 2/3 of those objects are probable H II regions, but 1/3 might be SNRs.  Overall, we conclude that the [S~II] criterion is a conservative way to identify SNRs. 

\subsubsection{Morphology}

At the distance of M31, 1\arcsec\/ corresponds to about 3.7 pc, so most SNRs are extended.  In Galactic nebulae, shock waves are distinguished by filamentary morphology, because the emission arises in a thin sheet behind the shock.  In extragalactic SNRs, it is not possible to resolve the filaments, but in principle the emission should take the form of a limb-brightened ring.  In most cases, however, the interaction of the SNR shock with pre-existing clouds leads to a partial ring or to a more complex morphology.  The nebula may be barely resolved, it may be irregular or diffuse, or it may be superposed on an H II region.  \citet{lee14} classified their SNR candidates as complete or partial shells, center-filled shells or diffuse (see their Table 3).  The objects they classified as complete shells satisfy the other criteria for SNRs, suggesting that morphology can be a helpful consideration.  The difficulties are that the classification is subjective and that the morphologies of known SNRs are very diverse. Overall, a shell morphology can provide some support for identification of a nebula as an SNR, but the absence of a clear ring does not indicate an H II region.  At present, the morphology is a subjective criterion.


\section{Additional SNRs \label{others}}

Our  high and low resolution spectoscopic observations included a large number of M31 objects not in the LL14 SNR list but identified as having strong  H$\alpha$, and initially thought to be H~II regions, PNe or symbiotic stars.  Also part of the data set are over 100 of the objects classified as SNR by others prior to LL14, but ruled out by LL14 based in their imaging analysis. These objects are all well-distributed in the star forming areas of M31 (and elsewhere as well for the PNe). The spectra were analyzed in the same manner as the LL14 SNR candidates, with the general result as described above that about 2700 objects have detectable [S~II]/H$\alpha$ \ measurements.  Several hundred have [S~II]/H$\alpha  > 0.4$, but the majority of those have  H$\alpha$ FWHM  less than 50 \kms (though measured in the low-resolution data) and are likely either the outer regions of large HII regions, or the Diffuse Interstellar Gas (DIG).  As well,  many of those with high line widths are simply a different part of one of the known LL14 SNRs that we have already observed (and are farther away from their centers than 10\arcsec). Thus despite the large number of objects available for consideration, there are only four new candidate SNRs, listed in Table \ref{other_snr}, along with the young SNR WB92-26 reported in \cite{caldwell2023}.  CRL-1 was part of the M31 VLA radio catalog by Toomey et al. (in prep.), which found this object as a radio source just outside of the \cite{AZ} catalog areal coverage though it does appear in the \cite{massey} \ha \ images, and of course CRL-2 comes from the \cite{AZ} sample. CRL-3 and CRL-4 were part of the \cite{miko14} survey for symbiotic stars, though these two were not symbiotics. One of these objects, CRL-3 = M95-3-88, was previously ruled out by LL14, but is indeed likely an SNR. The rest of the many objects previously declared not to be SNR by LL14 remain so based on the spectra. Diameters were measured by us on the \cite{massey} \ha \ images.

We further note that all four of the SNR listed in Table 1 of \cite{caldwell2023} as a comparison to WB92-26 under different names were indeed contained in the LL14 list, not just the two so-identified in the former paper.


\begin{deluxetable}{llrrrrrrl}
\tablecolumns{9}
\tablewidth{0pc}
\tablecaption{Additional SNRs, not in LL14}
\tablehead{\colhead{Object}   &\colhead{Alt} 
&\colhead{ra}
&\colhead{dec}
&\colhead{[S II]/H$\alpha $}
&\colhead{FWHM}
&\colhead{FW$_{20}$}
&\colhead{Diam}
&\colhead{Comment}\\
\colhead{} &\colhead{} &\multicolumn{2}{c}{J2000} &\colhead{} &\colhead{\kms} &\colhead{\kms}  &\colhead{\arcsec} &\colhead{}\\
 }
\startdata
 CRLL-1&K307-2, B-1440 &9.97392 & 40.92826 &1.03&163\tablenotemark{a}& 215&7&old SNR\\
 CRLL-2&AMB-585 &10.09420 & 40.48469 &0.71&90\tablenotemark{a}& 180 &5&\\
WB92-26    &AMB-1607  & 10.58513 & 41.44464 & 2.69 & 850\tablenotemark{b} & 2600  &2&young SNR\\
CRLL-3 & s279058, M95-3-88&11.15550 & 41.86685 &0.47&89\tablenotemark{a}& 125 &8& \\
CRLL-4 & s195399 &11.31417 & 41.65723 &0.46&64\tablenotemark{a}& 98 &$>5$&\\
 \enddata
\tablenotetext{a}{measured with high-resolution data.}
\tablenotetext{b}{measured with low-resolution data.}
\label{other_snr}
\end{deluxetable}

\section{Discussion \label{discuss}}





The two main questions for any survey of SNRs are 1) how confident can we be in identifying nebulae as SNRs using different criteria?  and 2) how complete is the sample? We consider those questions next.

\subsection{Fractions of incorrect identifications}

The line width measurements generally support the idea that nearly all of the nebulae having [S~II]/H$\alpha$\/ ratios larger than 0.4 are SNRs.  On the other hand, some of the candidates showing [S~II]/H$\alpha ~<~0.4$ have FW$_{20}$ line widths above 50 \kms\/ (see Figure~\ref{fig:s2ratio_fwhm}).   However, there is considerable scatter between the ratios given by \citet{lee14} and those measured from the Hectospec spectra (see Figure~\ref{fig_s2ratio_compare}), and line ratios measured from spectra should be preferred over those measured from narrow band images.  There is no sharp dividing line at [S~II]/H$\alpha$ = 0.4, however, and 5 of the \citet{lee14} candidates having ratios above 0.4 have very narrow line widths and are probably not SNRs. 

As noted above, there is no sharp dividing line in the distribution of [S~II]/H$\alpha$\ ratios at 0.4.  Among the objects we observed, 14 had [S~II]/H$\alpha$\ ratios between 0.3 and 0.4.  Of those, 7 had narrow line profiles indicative of H II regions, while 6 showed FW$_{20}$ broad enough to qualify as SNRs.  Three of those 6 are 2-component profiles, with narrow FWHM and wider wings, indicating a superposition of shocks and photoionized gas.  Eleven candidates whose [S~II] ratio was below 0.4 showed a line widths above 50 \kms, but they were objects where either the [S~II] ratio was above 0.3 or FW$_{20}$ was just barely above 50 \kms.  On the other side, there were 11 candidates with [S~II]/H$\alpha$\ ratios between 0.4 and 0.5, of which 2 had line widths indicating H II regions, and 2 were ambiguous cases, in that both the [S~II] criterion and the line width criterion were close to the cutoff values.

Overall, if two out of three criteria ([S~II], line width and X-ray/radio/UV emission) are satisfied, we can be quite confident of a supernova remnant identification.  In most of the cases that satisfy only two criteria, we do not have a measurement for the third.  There is a great deal of overlap between the [S~II] and line width criteria, and most of the cases where the two disagree are marginal, in that the [S~II] ratio is close to 0.4, the line width is near the cutoff or both.  In those cases, another criterion such as [O I] intensity, radio or UV detection might resolve the issue, but in general better optical and X-ray observations are needed.  If a galaxy can be covered with IFU observations, they will provide more accurate [S~II]/H$\alpha$\ ratios, along with [O I] intensities, morphological information and line profile information, which is an enormous improvement over any single criterion \citep{li2024, duarte-puertas24}.

\subsection{Level of incompleteness}

It is also important to keep in mind that we are systematically missing specific kinds of SNRs.  The star formation rate of 0.35-0.4 M$_\sun $ yr$^{-1}$ from \citet{m31_sfr} and supernova rate of \citet{li11} lead to an expectation of 320 to 700 SNRs in M31.  \citet{lee14} listed 156 candidate SNRs, of which we have found that 33 are probably not SNRs.  One additional SNR was identified by \citet{caldwell2023} and one by \citet{stiele11}, and four new candidates are listed in Section~\ref{others}.  Thus the observed SNRs in M31 account for only about 20-40\% of those expected for its star formation rate.  Many are missing, and the searches are biased against some specific SN types. 

Type Ia SNRs are likely to occur in low density regions of the ISM. Young Type Ia SNs may produce Balmer-line filaments, as in SN1006 and Tycho's SNR, but those are very faint.  Moreover, they would be excluded by the [S~II]/H$\alpha$ criterion.  They would appear in X-rays and radio, but they tend not to be very bright:  $L_X~\simeq~10^{34}~\rm erg~s^{-1}$ for SN1006, and its radio surface brightness is only $2\times 10^{-21}~W~m^{-2}~Hz^{-1}~sr^{-1}$.  Similarly, SNe that occur in a galactic halo or an elliptical galaxy encounter low density gas, and they are therefore faint.  A number of very large, faint SNRs in the Galactic halo have been discovered recently \citep{fesen24}, and they would not be detectable in other galaxies.

Many core collapse SNe occur in the star formation regions where the progenitors are born, and therefore in the superbubbles created by stellar winds and earlier SNe. \citet{lingenfelter19} suggests that 80\% of the CC SNe occur in superbubbles, which means that they would be in low density gas and therefore hard to detect.  Some of the class C objects that \citet{duarte-puertas24} found in M33 and NGC 6822 could be superbubbles, but they would be counted as single SNRs.

Finally, some objects are simply rare.  There is nothing similar to the Crab Nebula, though B0540-69 in the LMC is somewhat similar.  There are a handful of O-rich remnants related to Cas A, such as N132D in the LMC, but few O-rich remnants have been found in other galaxies. N-rich remnants are also quite rare, perhaps because surveys based upon narrow band images give artificially low [S~II]/H$\alpha$ ratios.  Therefore, WB92-26 was apparently unique \citep{caldwell2023} in its very high nitrogen overabundance as well as its high velocities.  Very recently \citet{kravtsov24} selected objects with very wide line profiles in the MUSE data cubes from the PHANGs survey, and they found 6 O-rich SNRs and one N-rich SNR similar to WB92-26. 

In the past, the number of SNRs that are contained in SNR catalogs has been used to estimate SN rates, but this is clearly gives only a lower limit.  To the extent that we know the star formation rate and the fraction of stars that become supernovae fairly accurately for the SFR, the numbers of SNRs tells us more about the galactic environments in which we can detect SNRs.

\section{A few interesting SNR \label{few}}

\begin{figure}[ht]
\includegraphics[width=8.0in]{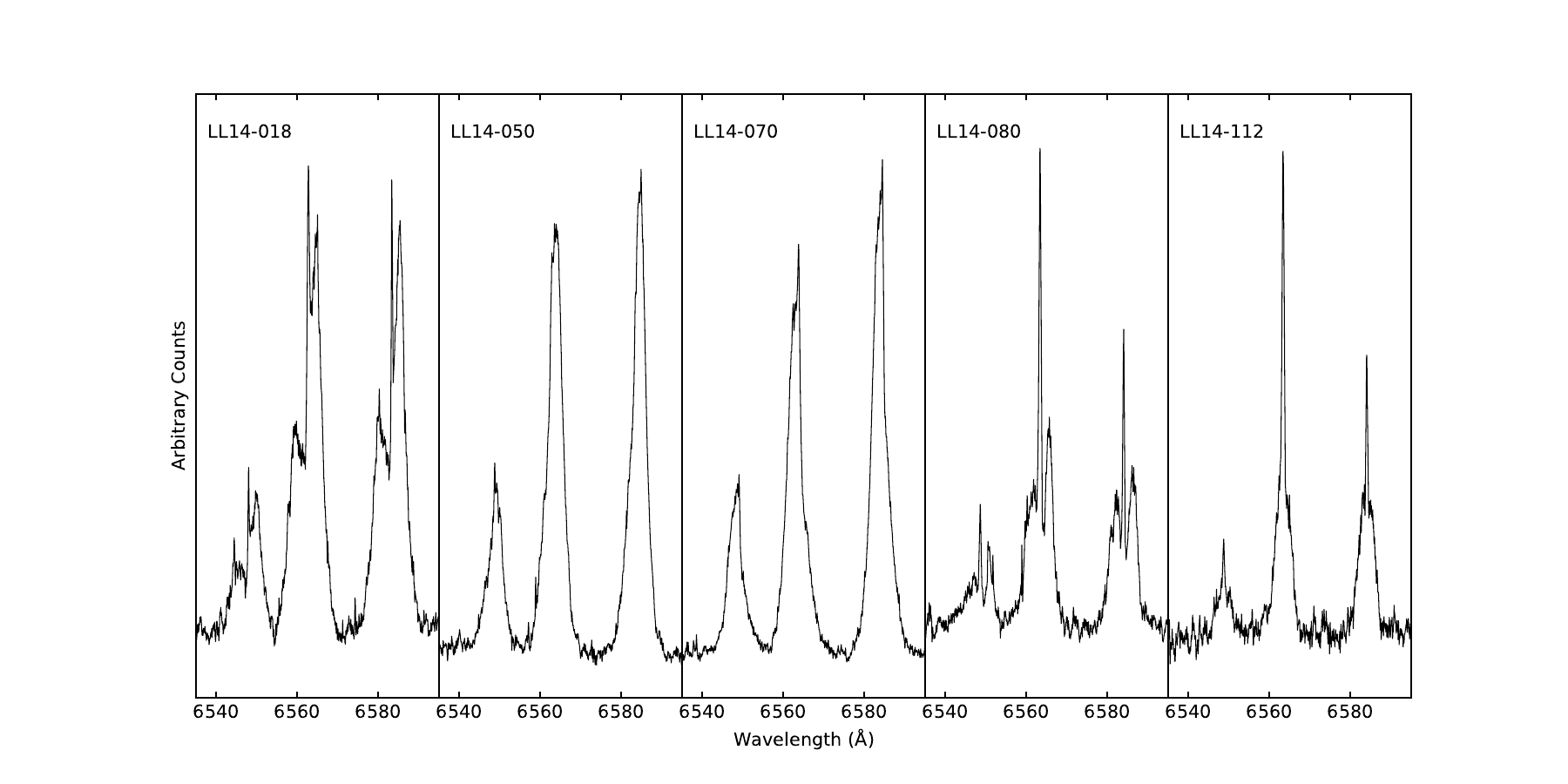} 
\caption{High resolution spectra showing \ha \ and the [N~II] lines of five SNR which have complex kinematics. Specifically, LL14-018 has a double profile along with a narrow spike.
LL14-050 and LL14-70 have broad wings. LL14-080  has a very broad profile with complex morphology.
LL24-112 has a wide profile along with a narrow spike. All five of these are detected in the X-ray and radio.}

\label{fig:interesting}
\end{figure}

Aside of the object LL14-124 already briefly described in section \ref{reduction}, there are a few more objects with interesting high resolution spectra we show here as a prelude to more detailed analysis or observation, see Fig. \ref{fig:interesting}. Images of these five as well as all the SNR and candidates are available at the M31 Hectospec web site \url{https://lweb.cfa.harvard.edu/oir/eg/m31clusters/M31_Hectospec.html} 




\section{Summary and Conclusions \label{summary}}


We have conducted a large high- and low-resolution spectroscopic survey of SNRs and SNR candidates in M31, based on the objects suggested to be SNRs by \cite{lee14}. Until now, the SNR population of M31 has been much less well studied than that of its neighbor M33. Our main conclusions are as follows:

\begin{itemize}  

\item
The overlap between the [S~II]/ \ha \ and FW$_{20}~>~50$ \kmss criteria is very high. Out of our sample of 152, there are 85 objects that satisfy both criteria.  Most of the 29 that appear to satisfy only one are objects for which we have only a high- or low-resolution spectrum, and it is likely that they would satisfy both criteria if we had both measurements. The exceptions are marginal cases with [S~II] ratios close to 0.4 and FW$_{20}$ close to 50 \kms.  The exceptions include 6 candidates just below the [S~II] ratio cutoff whose line widths indicate probably shock excitation and 2 candidates that satisfy the [S~II] criterion, but whose line widths indicate photoionized gas.

\item
A large fraction of these objects do satisfy the [S~II]/\ha\ criterion that has traditionally been used to certify a nebula as a SNR, validating the results based on earlier imaging studies.  However, the ratios reported by \citet{lee14} do not always agree with those measured from Hectospec spectra.  The objects whose spectra satisfy this criterion are almost certainly SNRs, though 2 or 3 have narrow line profiles that suggest photoionized rather than shock-excited gas.  

\item
Generally speaking, if an object satisfies the [S~II]/\ha\  criterion it also satisfies criteria based on other line ratios, such as [OI]/\ha, as is expected from shock models. We find that among the objects for which we have both high- and low-resolution spectra (130), 88 lie above the [S~II]/H$\alpha$ = 0.4 line, and SNR identifications of 85 of those are corroborated by line widths FW$_{20} >$ 50 \kmss and 2 more are marginal, with FW$_{20} \sim$ 50 \kms.  Thus the [S~II] ratio is a conservative criterion.  There are 27 objects above the [S~II]/H$\alpha$ = 0.4 line, but below the curve on the BPT diagram that \citet{kewley06} propose as the separation between shocked and photoionized gas.  We find that 85\% of these have broad enough lines to qualify as shocks, and others are marginal cases, so that we believe that the [S~II] criterion is more reliable than the BPT criterion.

\item
Further support for SNR identification of about 50 of the objects is provided by X-ray, UV or radio detections. However, these should be considered tentative. In particular, UV detections are less common and have been less well studied.  Radio detections support some of the SNR identifications, but fewer SNRs are available from radio surveys.  For these observations at other wavelengths, it is important to have radio spectral indices, X-ray hardness ratios and perhaps UV colors to avoid confusion with H II regions or background AGN.  

\end{itemize}

More studies of the SNRs in M31 are needed, particularly at radio wavelengths.  When feasible, IFU observations offer great advantages.

\begin{acknowledgments}
We wish to express our gratitude to the MMT staff who carried out the observations.
MGL was supported by the National Research Foundation grant funded by the Korean Government (RS-2024-00340832)
\end{acknowledgments}

%

\vspace{5mm}
\facilities{MMT(Hectochelle, Hectospec)}

\software{
IRAF \citep{iraf}
          }
          
\bibliography{bib}{}
\bibliographystyle{aasjournal}

\clearpage

\end{document}